\documentclass{gGAF2e}

\def\ga{\mathrel{\mathchoice {\vcenter{\offinterlineskip\halign{\hfil
$\displaystyle##$\hfil\cr>\cr\sim\cr}}}
{\vcenter{\offinterlineskip\halign{\hfil$\textstyle##$\hfil\cr
>\cr\sim\cr}}}
{\vcenter{\offinterlineskip\halign{\hfil$\scriptstyle##$\hfil\cr
>\cr\sim\cr}}}
{\vcenter{\offinterlineskip\halign{\hfil$\scriptscriptstyle##$\hfil\cr
>\cr\sim\cr}}}}}

\def\la{\mathrel{\mathchoice {\vcenter{\offinterlineskip\halign{\hfil
$\displaystyle##$\hfil\cr<\cr\sim\cr}}}
{\vcenter{\offinterlineskip\halign{\hfil$\textstyle##$\hfil\cr  
<\cr\sim\cr}}}
{\vcenter{\offinterlineskip\halign{\hfil$\scriptstyle##$\hfil\cr
<\cr\sim\cr}}}
{\vcenter{\offinterlineskip\halign{\hfil$\scriptscriptstyle##$\hfil\cr
<\cr\sim\cr}}}}}

\renewcommand{\vec}[1]{\mbox{\boldmath$#1$}}
\newcommand{\bH}{{\bf H}}
\newcommand{\bE}{{\bf E}}
\newcommand{\bj}{{\bf j}}
\newcommand{\bh}{{\bf h}}
\newcommand{\bn}{{\bf n}}
\newcommand{\ba}{{\bf a}}
\newcommand{\bB}{{\bf B}}
\newcommand{\bL}{{\bf L}}
\newcommand{\bbb}{{\bf b}}
\newcommand{\bX}{{\bf X}}
\newcommand{\Real}{\mathbb R}
\newcommand{\Omegac}{{\Omega_c}}
\newcommand{\Omegav}{{\Omega_v}}
\newcommand{\frontc}{{\Gamma_c}}
\newcommand{\front}{\Gamma}
\newcommand{\frontv}{{\Gamma_v}}
\newcommand{\ROT}{\nabla{\times}}  
\newcommand{\CROSS}{{\times}}
\newcommand{\DIV}{\nabla\!{\cdot}}   
\newcommand{\GRAD}{\nabla}           
\newcommand{\muc}{\mu^c}
\newcommand{\muv}{\mu^v}
\newcommand{\bHc}{\bH^c}
\def\Rm{R_{\text{\rm m}}}
\newcommand{\LAP}{{\Delta}}          
\newcommand{\bnc}{\bn^c}
\newcommand{\bnv}{\bn^v}
\newcommand{\SCAL}{{\cdot}}
\newcommand{\refp}[1]{(\ref{#1})}
\def\tildeHunv{{H^1_{\scriptscriptstyle\!\int\!=0} (\Omegav)}}
\def\Hrotc{\bH_{\text{\rm curl}}(\Omega_c)}
\def\Hunv{H^1(\Omegav)}

\begin{document}
\doi{10.1080/0309192YYxxxxxxxx}
\issn{1029-0419}
\issnp{0309-1929}
\jvol{00} \jnum{00} \jyear{2008} \jmonth{January}


\title{Generation of axisymmetric modes in cylindrical kinematic mean-field dynamos of VKS type}

\author{A. Giesecke$^{\rm a}$\thanks{$^\ast$Corresponding
    author. Email: a.giesecke@fzd.de \vspace{6pt}} , C. Nore$^{\rm
    b,c}$, F. Plunian$^{\rm d}$, R. Laguerre$^{\rm b,e}$,
  A. Ribeiro$^{\rm b}$, F. Stefani$^{\rm a}$, G. Gerbeth$^{\rm a}$,
  J. L\'eorat$^{\rm f}$, J.-L. Guermond$^{\rm b,g}$
  \\
  \vspace{6pt} $^{\rm a}${\em{Forschungszentrum Dresden-Rossendorf,
      Dresden, Germany}}
  \\
  \vspace{6pt} $^{\rm b}${\em{Laboratoire d'Informatique pour la
      M\'ecanique et les Sciences de l'Ing\'enieur, CNRS, BP 133,
      91403 Orsay cedex, France}}
  \\
  \vspace{6pt} $^{\rm c}${\em{Universit\'e Paris Sud 11, 91405 Orsay
      cedex, France et Institut Universitaire de France}}
  \\
  \vspace{6pt} $^{\rm d}${\em{Universit\'e Joseph Fourier, CNRS,
      Laboratoire de G\'eophysique Interne et de Tectonophysique,
      Grenoble, France}}
  \\
  \vspace{6pt} $^{\rm e}${\em{Universit\'e Libre de Bruxelles, CP.231,
      Boulevard du Triomphe, Brussels, 1050, Belgium}}
  \\
  \vspace{6pt} $^{\rm f}${\em{Luth, Observatoire de Paris-Meudon,
      place Janssen, 92195-Meudon, France}}
  \\
  \vspace{6pt} $^{\rm g}${\em{Department of Mathematics, Texas A\&M
      University 3368 TAMU, College Station, TX 77843, USA}} }

\maketitle

\begin{abstract}
  In an attempt to understand why the dominating magnetic
  field  observed in the von-K\'arm\'an-Sodium (VKS) dynamo experiment is
  axisymmetric, we investigate in the present paper the ability of
  mean field models to generate axisymmetric eigenmodes in
  cylindrical geometries. An $\alpha$-effect is added to the induction
  equation and we identify reasonable and necessary properties of the
  $\alpha$ distribution so that axisymmetric eigenmodes are
  generated. The parametric study is done with two different
  simulation codes. We find that simple distributions of $\alpha$-effect,
  either concentrated in the disk neighbourhood or occupying the bulk of the flow,
  require unrealistically large values of the parameter $\alpha$ to explain
  the VKS observations. 
\end{abstract}

\begin{keywords}
  Dynamo experiments; Induction equation; Kinematic simulations;
  alpha-effect;
\end{keywords}

\section{Introduction}
Dynamo action generated by a flow of conducting fluid is
the source of magnetic fields in astrophysical objects.
Homogenous dynamos also have been observed in three laboratory experiments
(Riga dynamo, \citealt{2000PhRvL..84.4365G}, Karlsruhe dynamo,
\citealt{2001PhFl...13..561S}, Cadarache von-K\'arm\'an-Sodium (VKS)
experiment, \citealt{2007PhRvL..98d4502M}).
The analysis of the dynamo effect benefits from complementary approaches of
scientific computing and experimental studies.
Indeed, experimental fluid dynamos offer an opportunity to test
numerical tools which can then be applied to natural dynamos. 
While the first two experimental dynamos produced results in agreement with
the predictions of numerical simulations the successful Cadarache
von-K\'arm\'an-Sodium experiment
brought interesting unexpected features: (i) dynamo action is observed
only with soft iron impellers and not with steel ones, (ii) the
axisymmetric component dominates (i.e. the azimuthal mode $(m=0)$) when
the dynamo action occurs.

Numerically simulating high permeability conductors embedded in
conducting fluids is a challenging task and requires the development of
new codes. We focus our attention in the present paper on a possible
numerical answer to the second question: how can an axisymmetric magnetic
field be generated?

The mode $(m=1)$ was observed as predicted from kinematic dynamo simulations
in the Riga experiment using axisymmetric
flows \citep{1980_gailitis,2004PhPl...11.2838G} 
and in the Karlsruhe experiment using an anisotropic $\alpha$-effect 
\citep{2002NPGeo...9..171R} 
or simulations based on the realistic configuration considerering the 52 spin
generators \citep{2002PhFl...14.4092T}.   
Since numerical simulations prior to the VKS experiment 
\citep{2003EPJB...33..469M, 2005PhFl...17k7104R, 2005physics..11149S}
also applied axisymmetric time averaged fluid flows, the mode $(m=1)$
was again predicted and utilized during the optimization process of the VKS
impellers.  (Recall that Cowling's theorem forbids the excitation of an
axisymmetric magnetic mode from an axisymmetric velocity field.)

Although the occurrence of the mode $(m=0)$ was a surprise, it does not
contradict physics but rather demonstrates that the axisymmetric flow
assumption is too simplistic. The counter-rotating impellers driving
the VKS flow are responsible for a relatively high turbulence level
compared to the first two dynamo experiments, and a large spectrum of
azimuthal modes must be taken into account. 
This may be done in various ways.
One can use the nonlinear approach \citep{2007PhRvE..75b6303B}, however
the Reynolds number which is achievable using direct numerical
simulations is thousand times smaller than the effective one, implying
that turbulence is still poorly described.
Alternatively, one can add some ad hoc non-axisymmetric flow modes in a kinematic 
code. Indeed, non-axisymmetric disturbances in terms of intermittent azimuthally drifting vortex
structures have been observed in water experiments by
\citet{these..marie} and \citet{2007PhRvL..99e4101D} but their influence on the dynamo process is
unknown. 
A third possibility is provided by the application of a mean field model where
the unresolved non-axisymmetric small scale fluctuations are parametrized by an
$\alpha$-effect \citep{1980mfmd.book.....K}.
Although the mean field
approach is somewhat controversial for large magnetic Reynolds numbers (see
e.g. \citealt{2006PhRvL..96c4503C,2008MNRAS.385L..15S}), it is numerically far less demanding
than direct numerical simulations and allows the exploration of the
parameter space more easily. This is the approach that we follow in the
present study since it receives support from the second order correlation
approximation (SOCA) as discussed
in section~\S~\ref{sec:alpha_model}.

Assuming scale separation, the mean field approach parametrizes
the induction action of unresolved small scale fluctuations via the $\alpha$-effect.
Our purpose is to determine whether an $\alpha$-model can produce
axisymmetric modes with realistic values of $\alpha$.  Using $\left<\cdot\right>$
to denote the averaging operation and primes for unresolved quantities,
the most simple $\alpha$-model states that
\begin{equation}
\left<\vec{u'} \times \vec{b'}\right>  = \alpha \vec{B},
\end{equation}
where $\vec{u'}$ denotes the fluctuating velocity field,
$\vec{b'}=\mu_0 \bh'$ the fluctuating magnetic flux density or
induction, $\bh'$ the fluctuating magnetic field, $\mu_0$ the vacuum
permeability, $\vec{B}$ the mean magnetic induction, and $\alpha$
the pseudo-tensor representing the $\alpha$-effect.  
The dynamo action is
analyzed by solving the kinematic mean field induction equation
\begin{equation}
  \frac{\partial\vec{B}}{\partial
    t}=\nabla\times\left(\vec{u}\times\vec{B}+\alpha\vec{B}-\eta\nabla\times\vec{B}\right) 
\label{eq::inductioneq}
\end{equation}
where $\vec{u}$ denotes a prescribed large scale velocity field and
$\eta$ the magnetic diffusivity ($\eta=1/\mu_0\sigma$ with $\sigma$
the electric conductivity).
In the kinematic approach the velocity field is given and any
back-reaction of the magnetic field on the flow by the Lorentz force is
ignored. 

In case of a homogenous, isotropic,
non-mirrorsymmetric small scale flow, the $\alpha$-tensor becomes
isotropic and reduces to a scalar.
The $\alpha$-effect provides an additional induction source by
generating a mean current parallel to the large scale field.
In general, the huge spatial extensions of astrophysical objects
ensure that even a small and localized $\alpha$-effect, potentially
supported by shear, is able to generate a large scale
field (see for example \citealt{2005LRSP....2....2C} for a review of current models of
the solar magnetic cycle).  
In the VKS experiment a possible source of the $\alpha$-effect is the
kinetic helicity caused by the shear between the outward driven
fluid flow trapped between adjacent impeller blades and the slower moving
fluid in the bulk of the container \citep{Petrelis07}.
Unfortunately, no helicity measurements are available for this
experiment and the magnitude and the spatial distribution of the
corresponding $\alpha$-effect are not known. Although sign changes can
occur within the domain, there are two limits for the $\alpha$
distribution. Either it is concentrated in the impeller region
\citep{2008PhRvL.101j4501L, 2008PhRvL.101u9902L}, or it is uniformly
distributed in the cylinder. We analyze these two extreme configurations
in the paper.

The present work attempts to identify essential properties of the
$\alpha$-effect that are necessary to generate an axisymmetric
magnetic field in VKS-like settings.  
In the VKS experiment, the critical
magnetic Reynolds number is $\rm{Rm}^{\rm{crit}} \approx 32$ using
soft iron impellers, where $\rm{Rm}$ is the magnetic Reynolds number defined
as usual as the
ratio between the stretching and the diffusive terms (see Eq.~\ref{eq::def_magReynolds} below). 
Kinematic simulations are applied close to the onset of dynamo action and are
used for the estimation of the magnitude of the $\alpha$-effect
necessary to produce the quoted dynamo threshold.
 
One important numerical difficulty consists of implementing
realistic boundary conditions on the magnetic field between the conducting
region and the vacuum. This is tackled in different ways in the two codes
described in the Appendix. One code is based on the coupling of a boundary element
method with a finite volume technique (FV/BEM, \citealt{2008giesecke_maghyd}), the other is based on the coupling of a
finite element method with a Fourier approximation in the azimuthal
direction (SFEMaNS, \citealt{GLLN09}). The two codes are validated by
comparing the results they give on identical problems.

The paper is organized as follows. In section~\ref{sec:axi_alpha_2},
we consider a cylinder with no mean flow, the dynamo action is caused
solely by an $\alpha^2$-effect. The simplified $\alpha$-tensor is
similar to the one used in the modeling of the Karlsruhe experiment.  We
compare periodic and non-periodic axial boundary conditions.  We validate
our two independent codes by observing that they give the same results
up to non-essential approximation errors.
In section~\ref{sec:VKS_alpha}, we model the geometry and the flow
pattern of the VKS experiment. The mean velocity field is obtained
from water experiments. The $\alpha$-effect is investigated and the
growth rates of the modes $(m=0)$ and $(m=1)$ are compared. Conclusions are
proposed in section~\ref{Sec:Conclusion}.

\section{Axisymmetric $\alpha^2$-dynamos in cylinders}
\label{sec:axi_alpha_2}
\subsection{$\alpha$-model}
We use cylindrical coordinates ($r, \, \theta, \, z$) throughout the paper.
We first focus our attention on a cylindrical mean field dynamo model with
vanishing mean flow $\vec{u}=0$.
The $\alpha$-effect is the only source of dynamo action and provides the necessary coupling
between the poloidal and toroidal fields which is essential to drive
the dynamo instability. 
The induction equation \eqref{eq::inductioneq} reduces to
\begin{equation}
\frac{\partial \vec{B}}{\partial
  t}=\nabla\times\left(\alpha\vec{B}-\eta\nabla\times\vec{B}\right).
\end{equation}

To bound the parameter space, we consider
a spatially constant but anisotropic
$\alpha$-effect that we assume to be nonzero 
only in the annulus $R_i \le r \le R_o$. We choose a simplified version of
the $\alpha$-effect modeling the forced helical flow realized in the
Karlsruhe experiment where  
$\alpha$ has the following tensor form:
\begin{equation}
\alpha=\left(\begin{array}{ccc}\alpha_0 & 0 & 0\\0&\alpha_0&0\\0&0&0_{ }\end{array}\right)
\label{eq:alpha_K}
\end{equation}
The motivation for this approach originates from
the modeling of turbulent flow structures in fast rotating spheres
and planetary bodies like the Earth.
Simplified spherical $\alpha^2$--dynamos with scalar, isotropic,
$\alpha$-tensor and zero mean flow are known to exhibit properties
that resemble essential characteristics of the Earth's magnetic field
like dipole dominance and reversal sequences.
However, more realistic spatial distributions of the $\alpha$-effect like
the anisotropic structure \eqref{eq:alpha_K} usually generate
equatorial dipole dynamos \citep{2005AN....326..693G} unless anisotropic
turbulent diffusivity is accounted for \citep{2004GApFD..98..225T}.

In the rest of section~\ref{sec:axi_alpha_2} we analyze the dynamo
action in a cylinder of radius $R$ and vertical extension $H$.  We
evaluate the influence of the geometric constraints of the container (in
terms of the aspect ratio $H/R$) and of the spatial distribution of
the $\alpha$-effect (homogenous or concentrated in the annulus $R_i
\le r \le R_o$) on the dynamo threshold.  We define the critical
dynamo number
\begin{equation}
C_{\alpha}^{\rm{crit}}=\displaystyle\frac{\alpha_0^{\rm{crit}}R}{\eta}
\end{equation}
where $\alpha_0^{\rm{crit}}$ is the value of $\alpha_0$ 
at which dynamo action occurs.

\subsection{Axisymmetric dynamos  with a uniform $\alpha$ distribution}

The aspect ratio of the Karlsruhe device is $H/R\approx 1$ according to the
overall size of the cylindrical container. 
When looking for periodic solutions
for magnetic fields, since $H$ would represent the half wavelength of a
transverse dipole, we have to consider periodic cylinders of aspect ratio
equal to two.
In this case, the $\alpha$-effect
is uniformly distributed in the cylinder. Using the SFEMaNS code, we
have verified that $C_{\alpha}^{\rm{crit}}=4.8$ for the mode
$(m=1)$. This result agrees with the Karlsruhe analytical modeling
from~\cite{2003PhRvE..68f6307A}.  The magnetic field observed in the
Karlsruhe experiment was indeed perpendicular to the cylinder
axis and described as an equatorial dipole. 
Note, that the critical dynamo number for the axisymmetric mode
$(C_{\alpha}^{\rm{crit}}=7.15)$ is larger than 
the one for the mode $(m=1)$ explaining that the mode ($m=0$) was not
observed. 

\begin{figure}
\centering\includegraphics[width=0.66\textwidth, bb=64 351 548 703,clip=true]{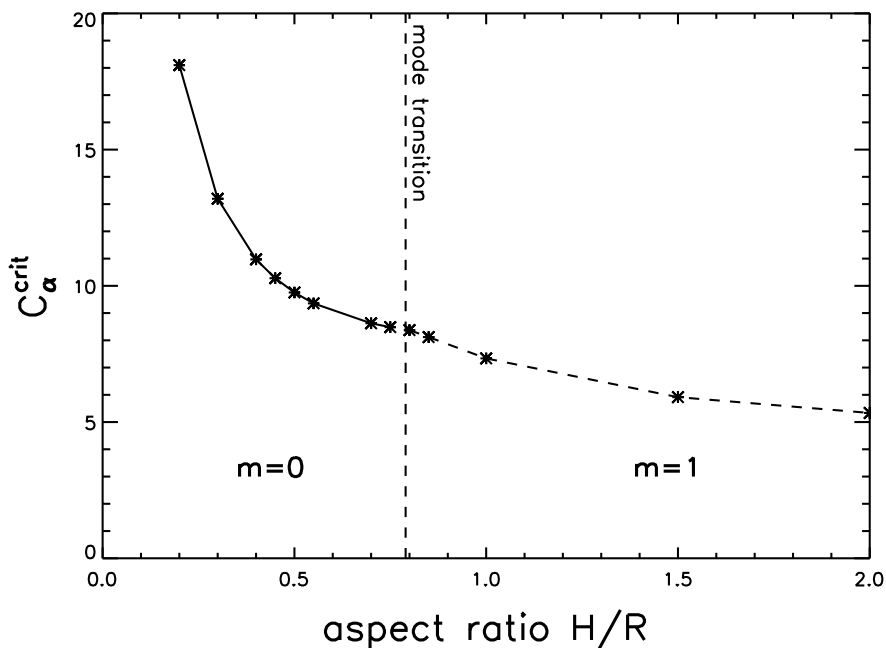}
\caption{Critical dynamo number in dependence of the aspect ratio $H/R$
  obtained from the FV/BEM algorithm. The
  transition between the modes $(m=0)$ and $(m=1)$ occurs at $H/R\approx 0.79$.}
\label{plot::calpha_krh}
\end{figure}

We now investigate whether the axisymmetric mode can become dominant
when the aspect ratio $H/R$ varies. We abandon the assumption of
periodicity along the $z$-axis, and we consider the finite cylindric
geometry. A uniform spatial distribution of $\alpha$ is still assumed.
We compute the critical dynamo number for various aspect ratios in
the range $[0.2; 2.0]$.  The results obtained with the hybrid FV/BEM
code are reported in Figure~\ref{plot::calpha_krh}. Computations done with the
SFEMaNS code give almost identical results.  These results have been shown
\citep{stefani_numsimexpdyn2009} to be similar to those obtained by \citet{2007GApFD.101..389A} applying the
integral equation approach \citep{2008JCoPh.227.8130X}.

The main conclusion that we can draw from these computations is that
the structure of the magnetic eigenmode passes from an equatorial
dipole $(m=1)$ to an axial dipole $(m=0)$ as the container geometry
passes from an elongated cylinder to a flat disk.  The change of
structure of the dominating mode is observed when the aspect ratio is
about $H/R \approx 0.79$.  If the aspect ratio of the Karlsruhe dynamo
($H/R\approx 0.84$ when the curved ended pipes are disregarded) 
had been chosen 10\% smaller, an axisymmetric
mode might have been produced. This would have enhanced the analogy of the model with
geomagnetism. Since the experiment is now
dismantled, further numerical work on this point is rather academic.

\begin{figure}
\begin{minipage}{0.48\textwidth}
\includegraphics[width=0.7\textwidth,bb=59 349 266 710,clip=true]{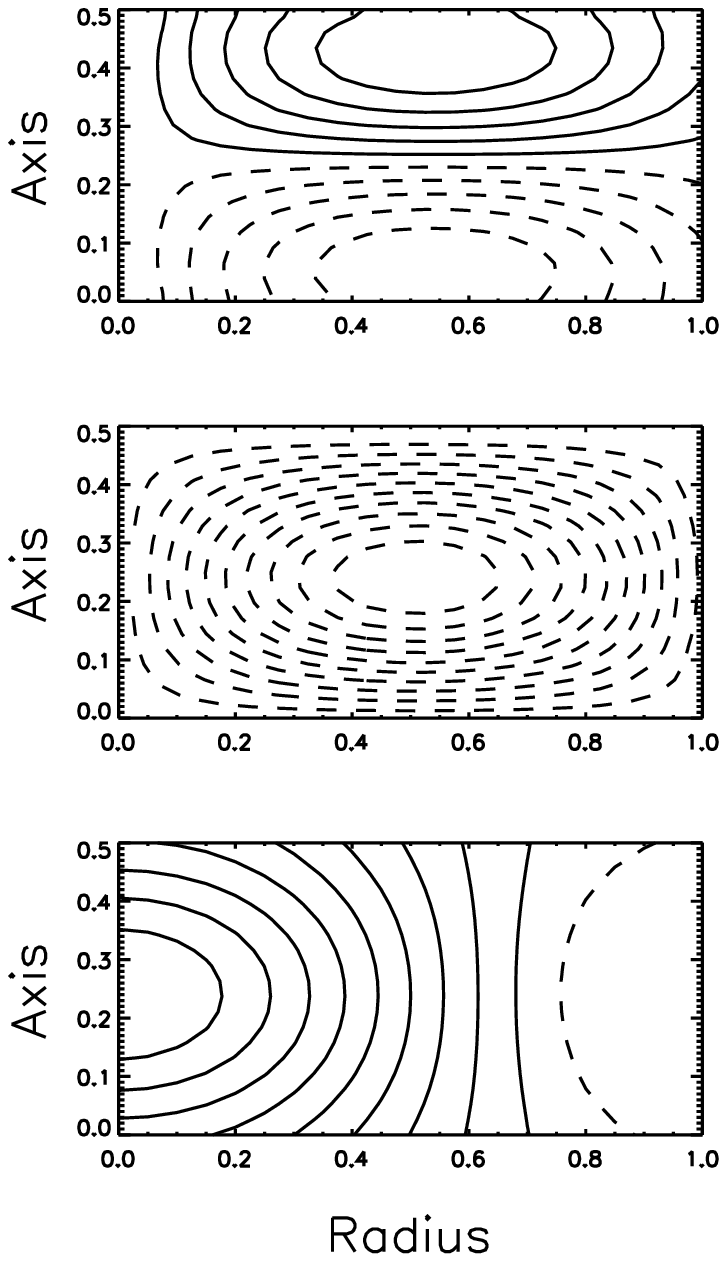}%
\end{minipage}
\begin{minipage}{0.52\textwidth}
\includegraphics[width=\textwidth,bb=45 340 552 624,clip=true]{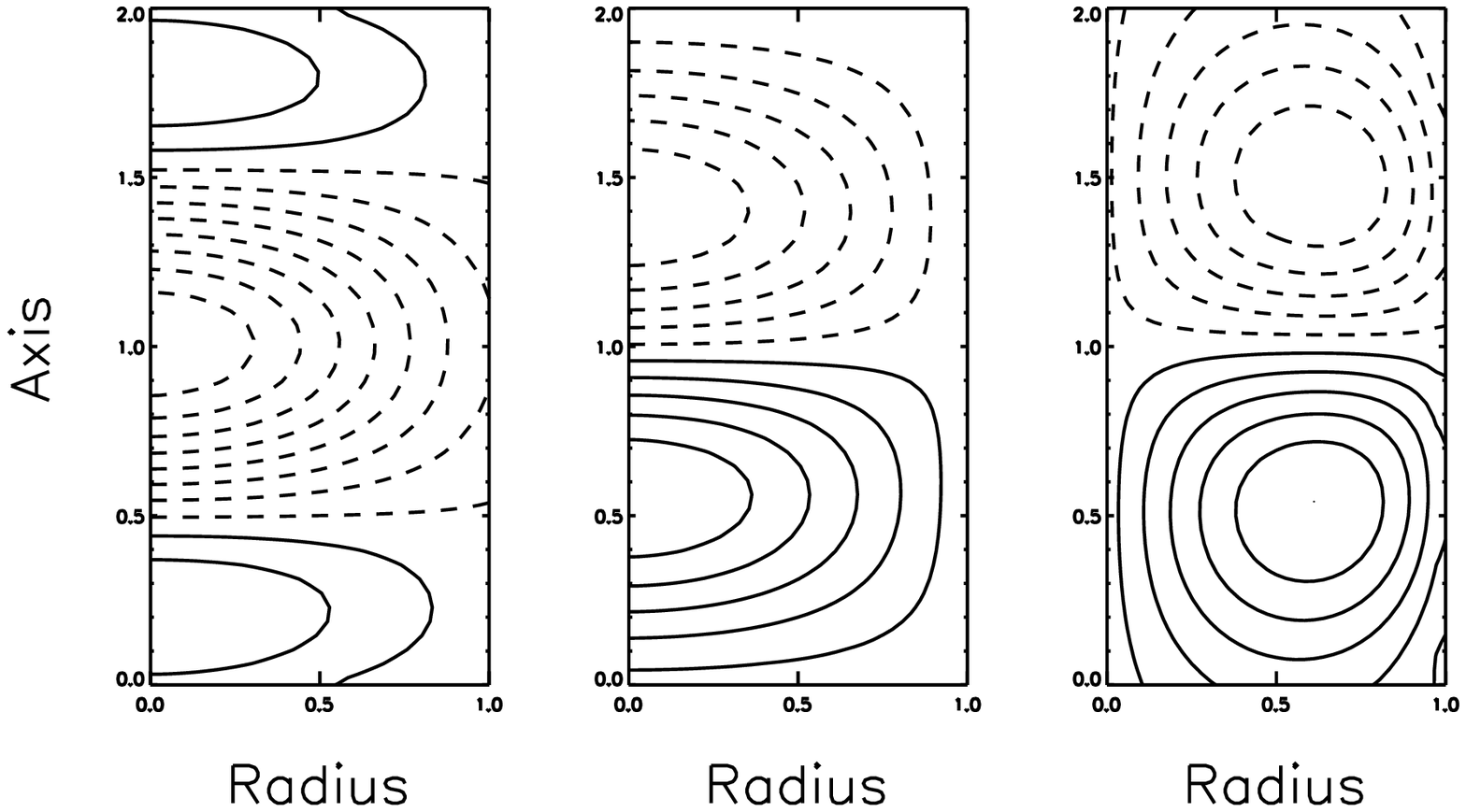}
\includegraphics[width=\textwidth,bb=45 340 552 624,clip=true]{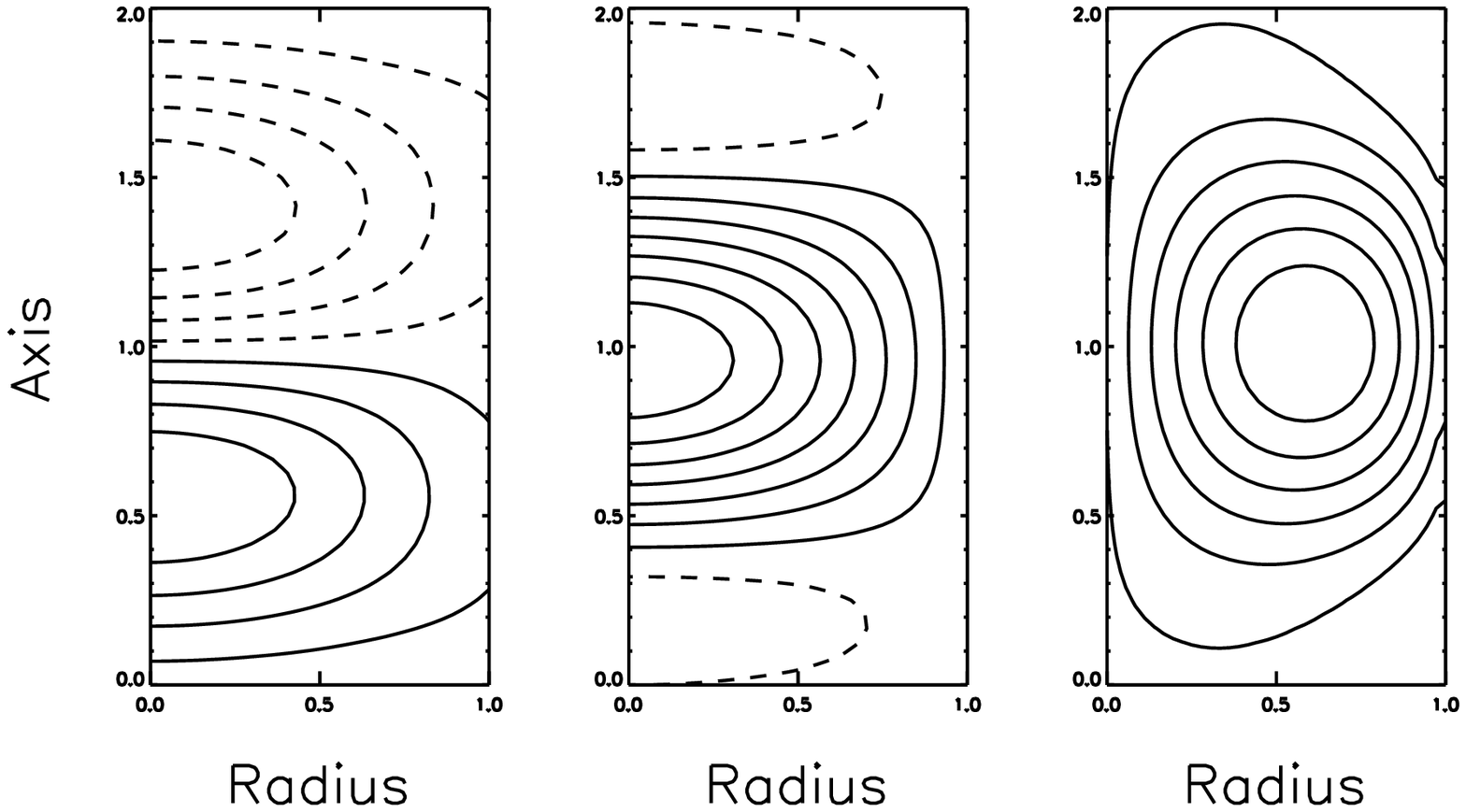}%
\end{minipage}
\caption{ Geometric structure of the magnetic field of the
  $\alpha^2$-dynamo in a finite cylinder with uniform
  $\alpha$-effect (obtained from FV/BEM approach). Left side: $H/R=0.5$, axisymmetric field at the
  marginal value $C_{\alpha}^{\rm{crit}}=9.8$ (from top to bottom:
  $B_r, B_{\theta}, B_z$). Right side: $H/R=2$, non-axisymmetric
  magnetic field in two meridional planes that differ by an angle of
  90 degrees at $C_{\alpha}^{\rm{crit}}=5.3$. (from left to right:
  $B_r$, $B_{\theta}$, $B_z$).  Solid curves denote positive values,
  dashed curves denote negative values.} %
\label{plot::axisym_field_struc_anisoa2_hr05}
\end{figure}

The eigenmodes associated with the modes $(m=0)$ for $H/R=0.5$ and
$(m=1)$ for $H/R=2$ are shown in
Fig.~\ref{plot::axisym_field_struc_anisoa2_hr05}.  The three panels on
the left-hand side show the components of the axisymmetric magnetic
field for the aspect ratio $H/R=0.5$ and the six panels on the
right-hand side show the eigenmode for $H/R=2$.  In the second case
the geometric structure of the non-axisymmetric magnetic field is
shown in two orthogonal meridional planes.
Note, that as the azimuthal wavenumber passes from $(m=1)$
(right) to $(m=0)$ (left), the typical scale in the axial direction
passes from $\sim 2$ for $H/R=2$ to less than 1
for $H/R=0.5$. When decreasing the aspect ratio, the geometric constraint
on the axisymmetric current increases $C_{\alpha}^{\rm{crit}}$ by a factor of $2$.

\subsection{Axisymmetric dynamos with  an annular  $\alpha$ distribution}

We keep the aspect ratio $H/R=2$ which produces a transverse field
when $\alpha$ is uniform, and we show now that axisymmetric dynamos can be
obtained when the distribution of $\alpha$ is annular,
i.e. $\alpha(r)\not=0$ if $r\in [R_i,R]$ and $\alpha(r)=0$ if $r\in
[0,R_i)$.  This study is inspired by unpublished analytical results of
\citet{azp..2009} in periodic cylinders, where a change of structure
of the dominating mode is observed for an annular distribution when
$R_i/R \approx 0.7$.

\begin{figure}
\centerline{
\epsfig{file=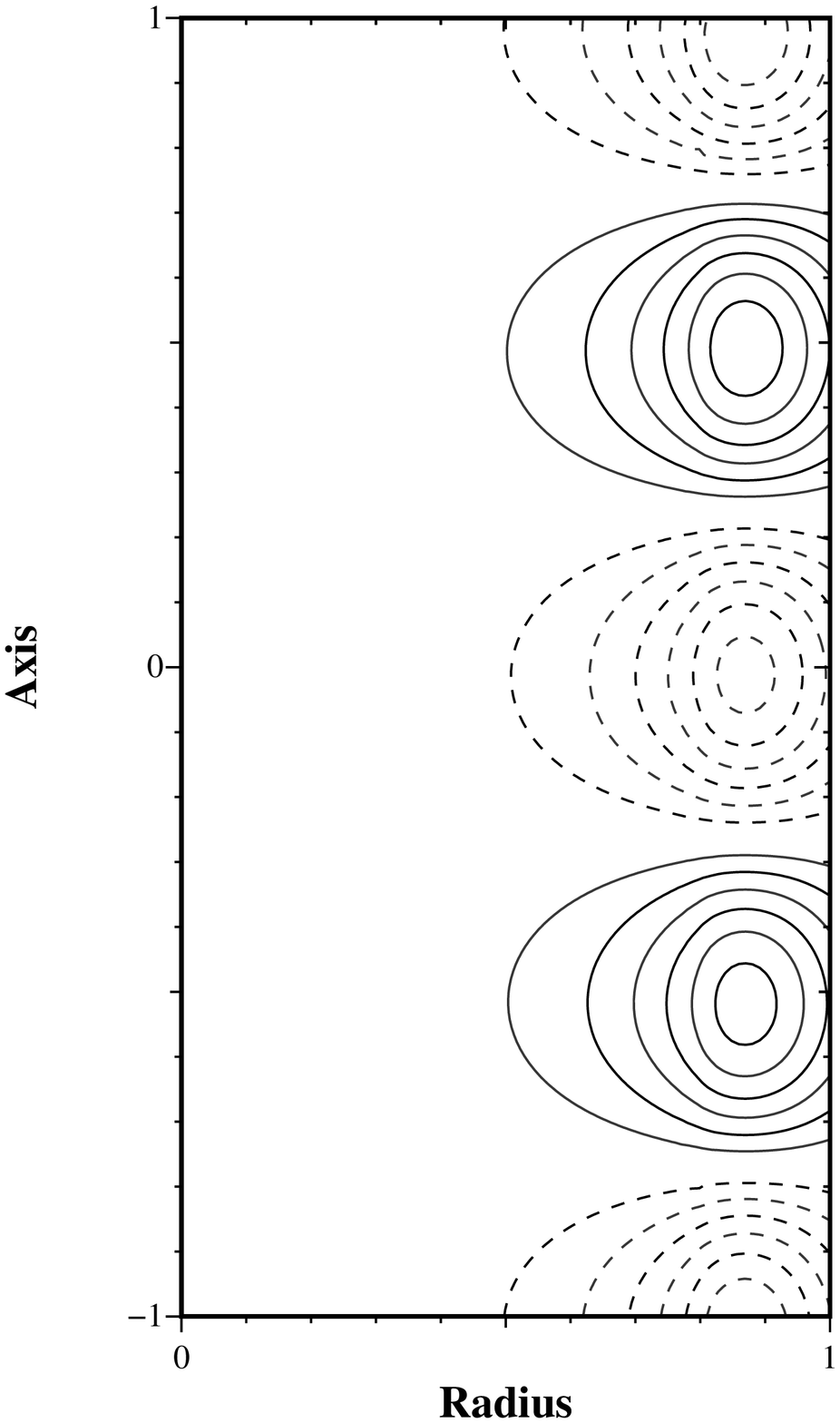,width=0.15\textwidth,bb=130 74 454 683, clip}
\epsfig{file=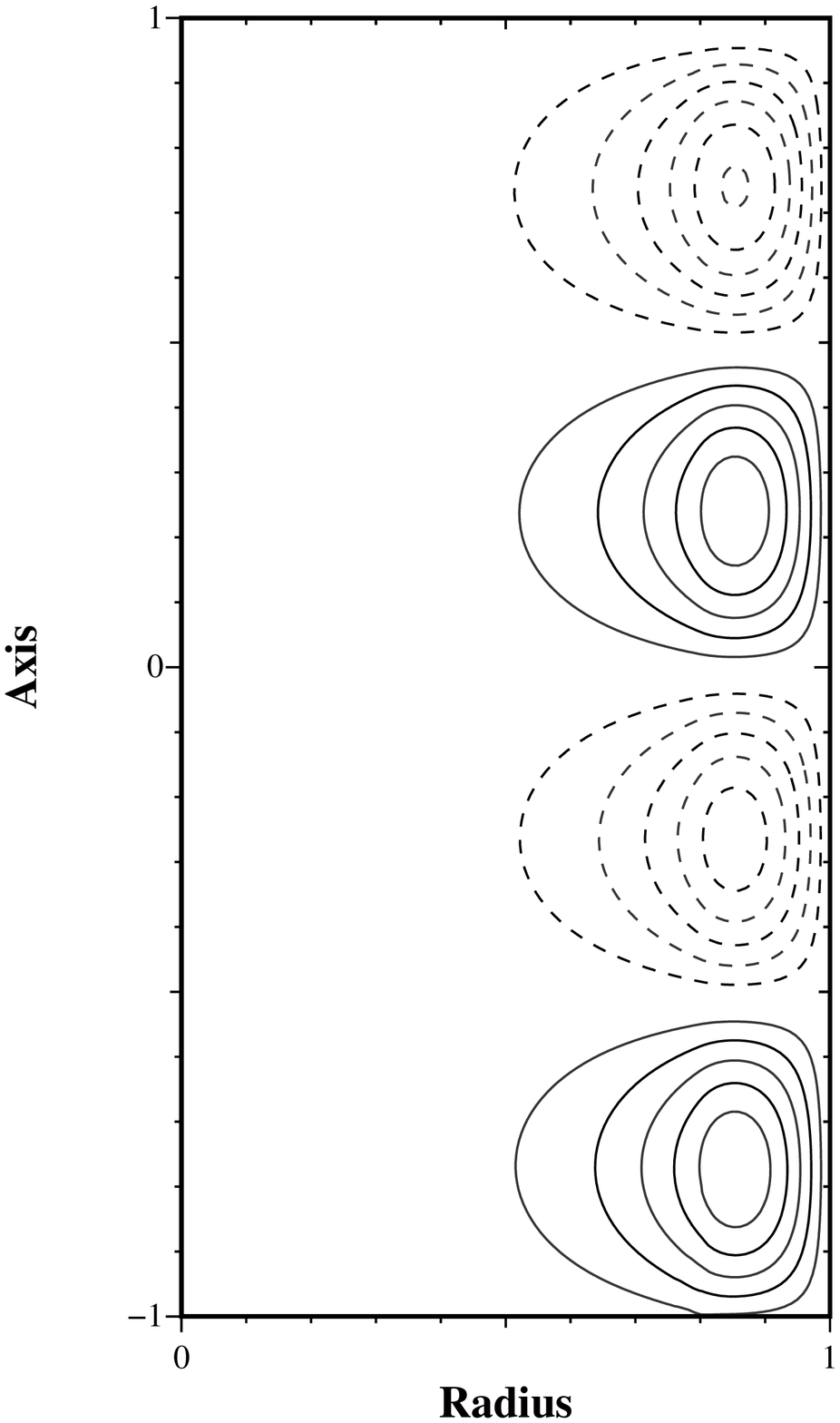,width=0.15\textwidth,bb=130 74 454 683, clip}
\epsfig{file=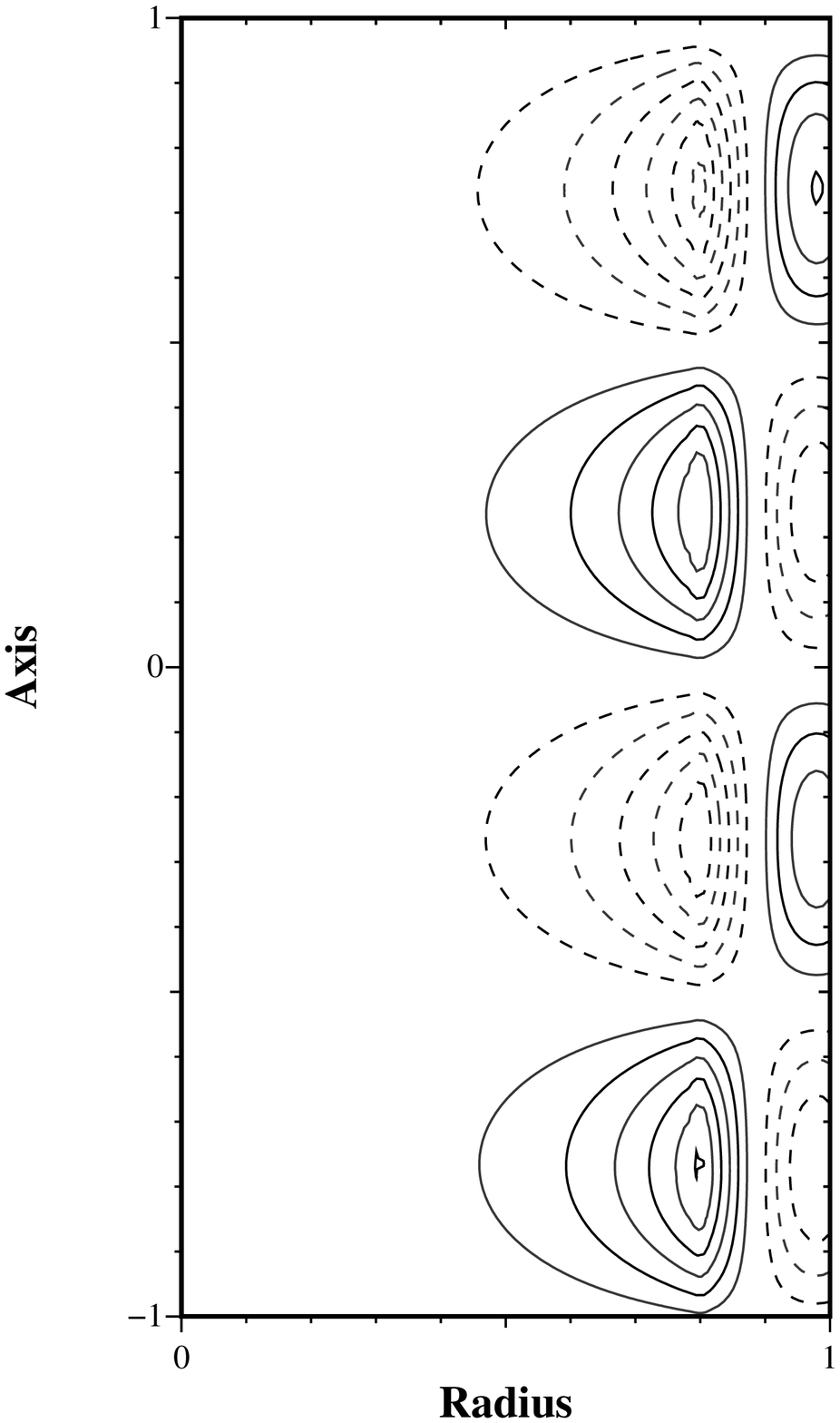,width=0.15\textwidth,bb=130 74 454 683, clip}
\hfil \hfil \hfil \hfil \hfil
\epsfig{file=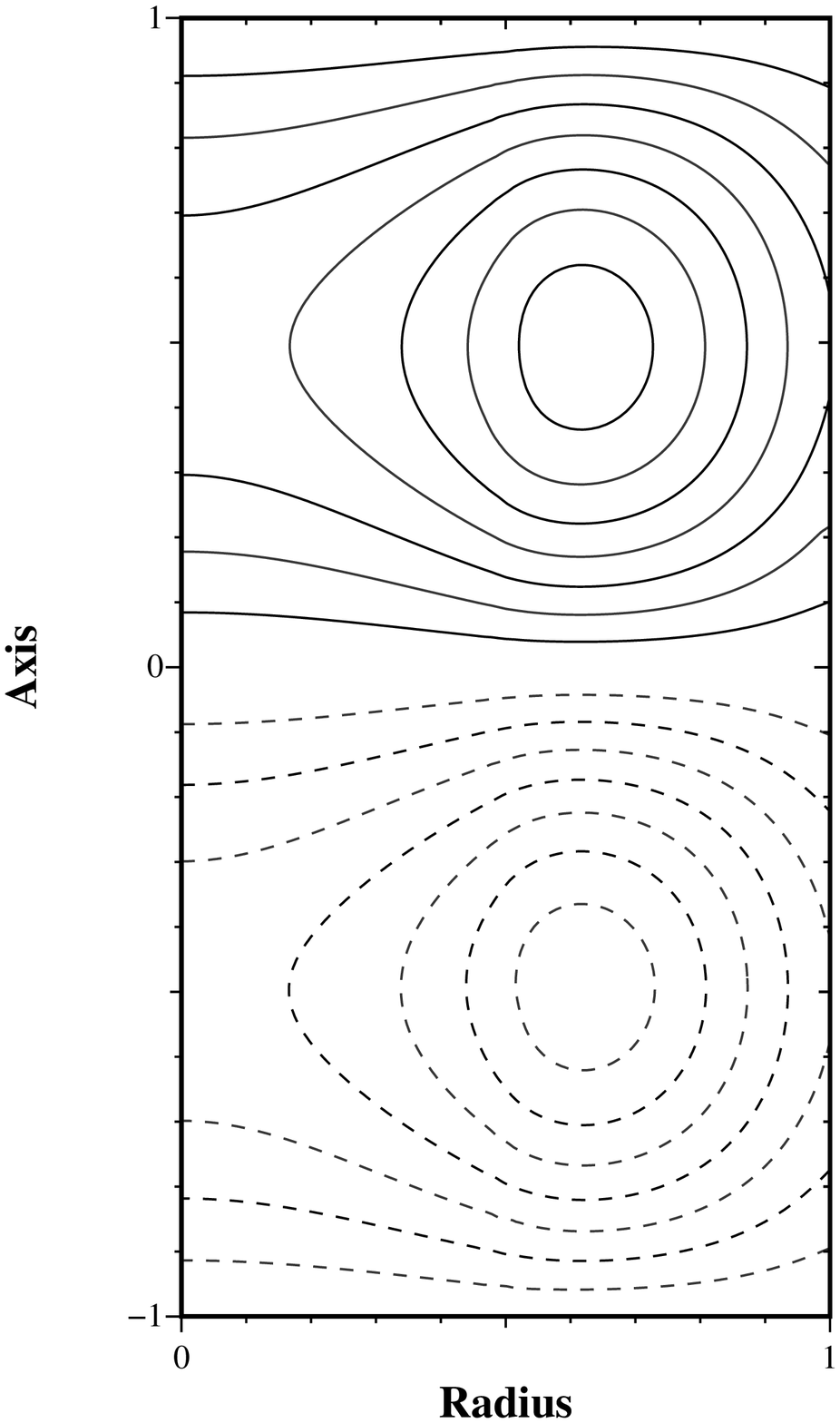,width=0.15\textwidth,bb=130 74 454 683, clip}
\epsfig{file=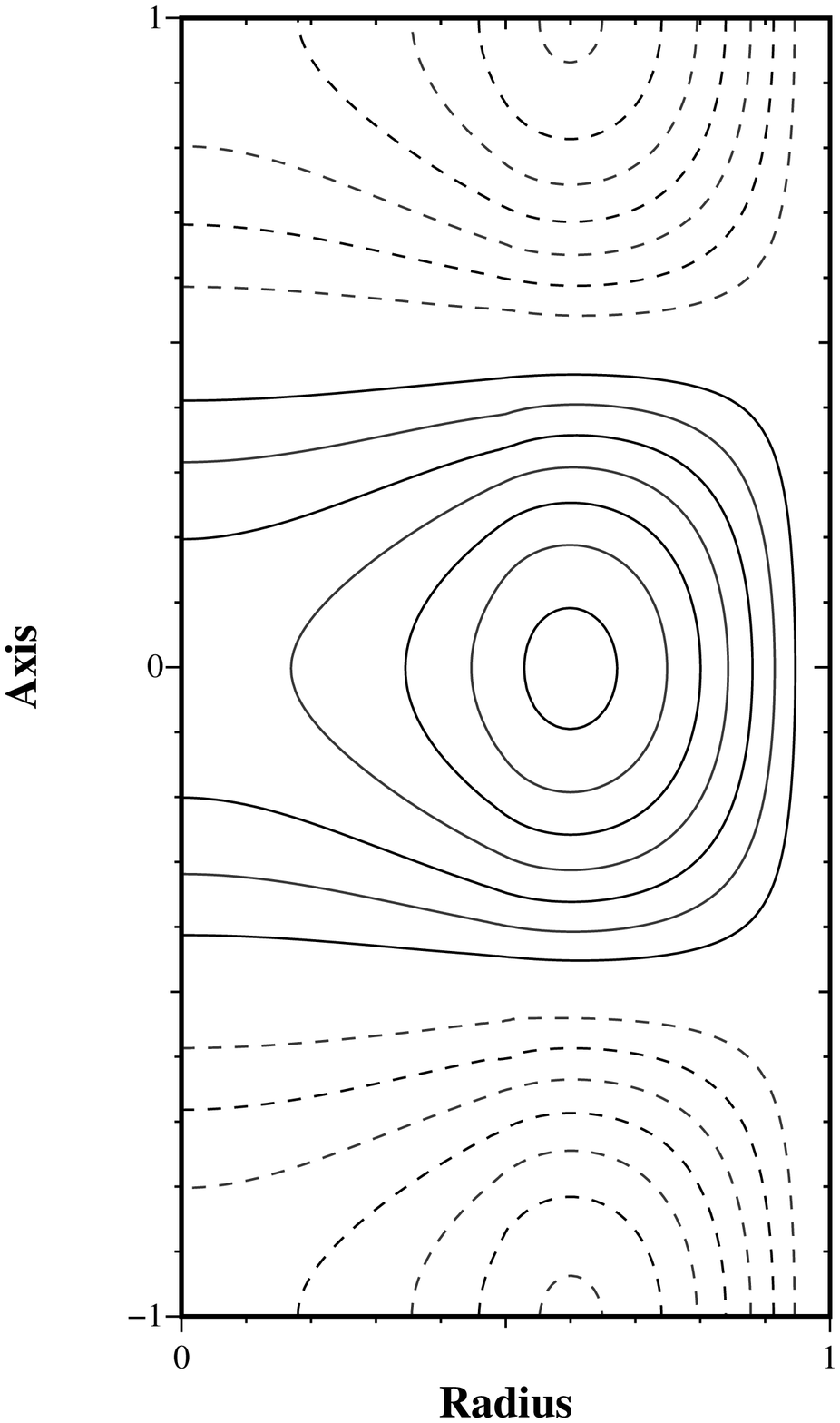,width=0.15\textwidth,bb=130 74 454 683, clip}
\epsfig{file=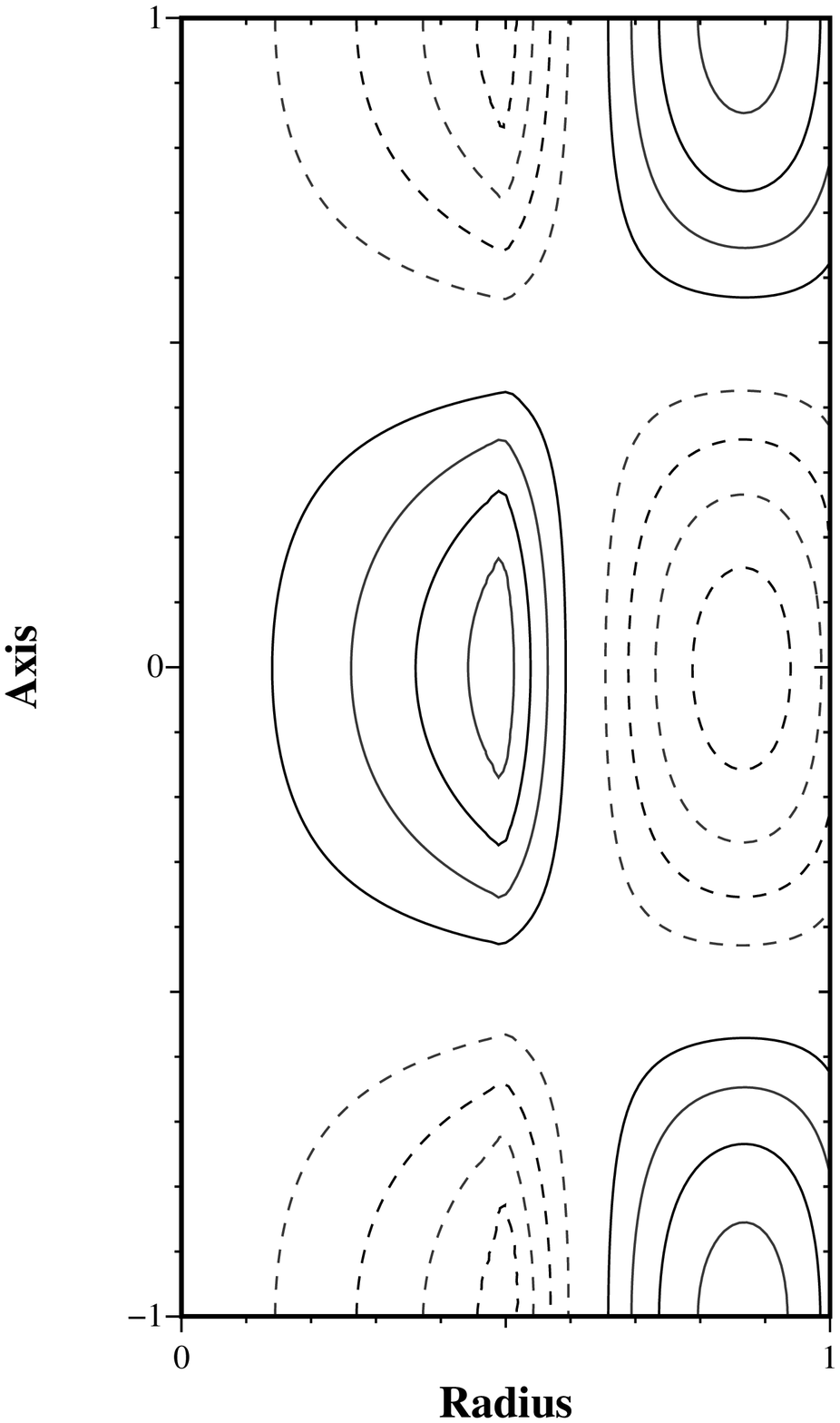,width=0.15\textwidth,bb=130 74 454 683, clip}
}
\centerline{(a) \hfil \hfil (b) \hfil \hfil (c) \hfil \hfil \hfil \hfil \hfil (a') \hfil \hfil (b') \hfil \hfil (c') \hfil}
\caption{Geometric structure of the magnetic field of the
  $\alpha^2$-dynamo in a periodic cylinder of aspect ratio $H/R=2$
  with an annular $\alpha$-effect: (a-c) $(m=0)$,
  $C_{\alpha}^{\rm{crit}}=20$ and $R_i/R=0.8$; (a'-c') $(m=1)$,
  $C_{\alpha}^{\rm{crit}}=8$ and $R_i/R=0.5$.  Represented are the
  radial (a,a'), azimuthal (b,b') and axial (c,c') components in the
  plane $\theta=0$. Data results from the SFEMaNS approach.}
\label{fig:EsR_perio}
\end{figure}

\begin{figure}
  \centerline{
    \epsfig{file=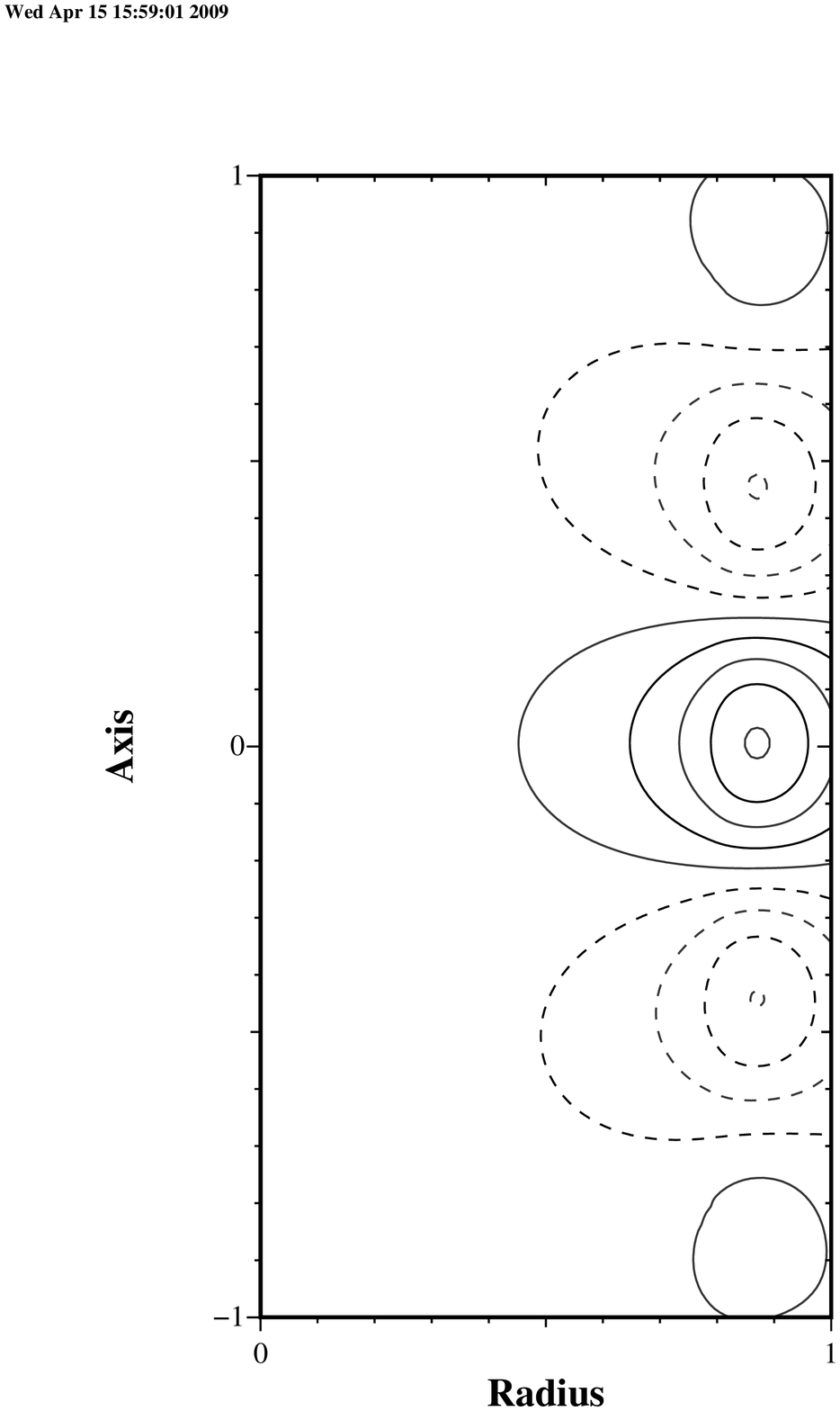,width=0.15\textwidth,bb=130
      76 451 683, clip}
    \epsfig{file=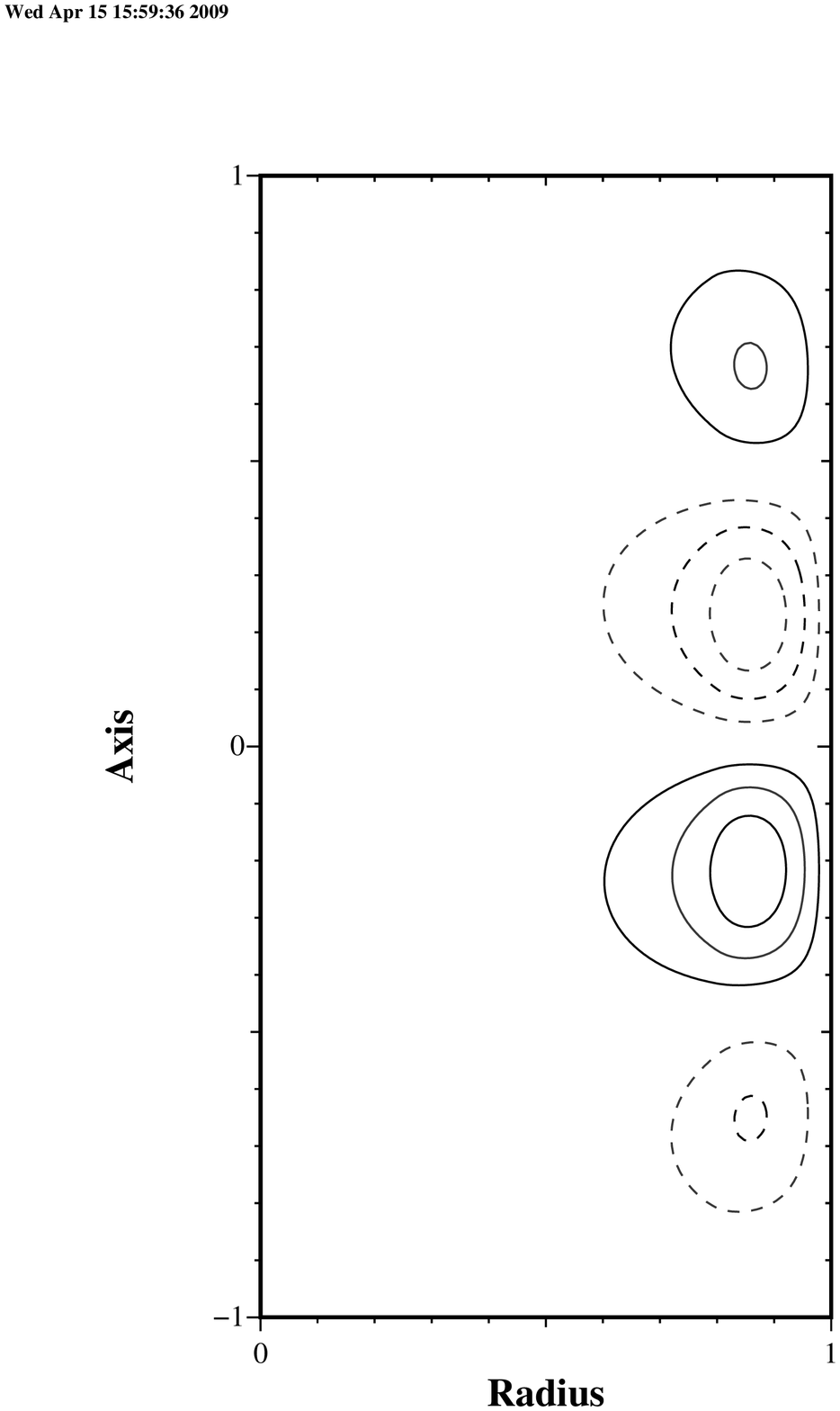,width=0.15\textwidth,bb=130
      76 451 683, clip}
    \epsfig{file=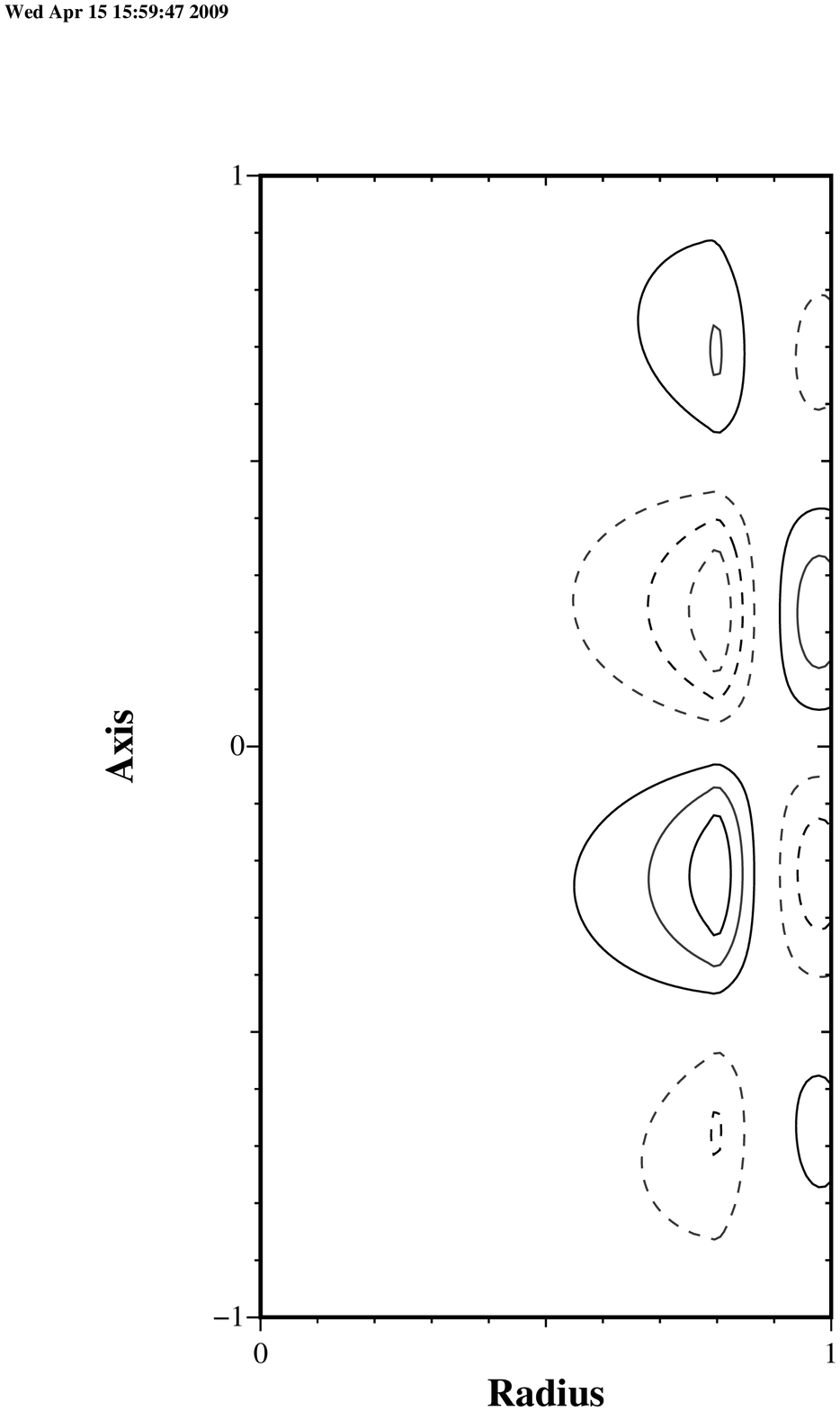,width=0.15\textwidth,bb=130
      76 451 683, clip} \hfil \hfil \hfil \hfil \hfil
    \epsfig{file=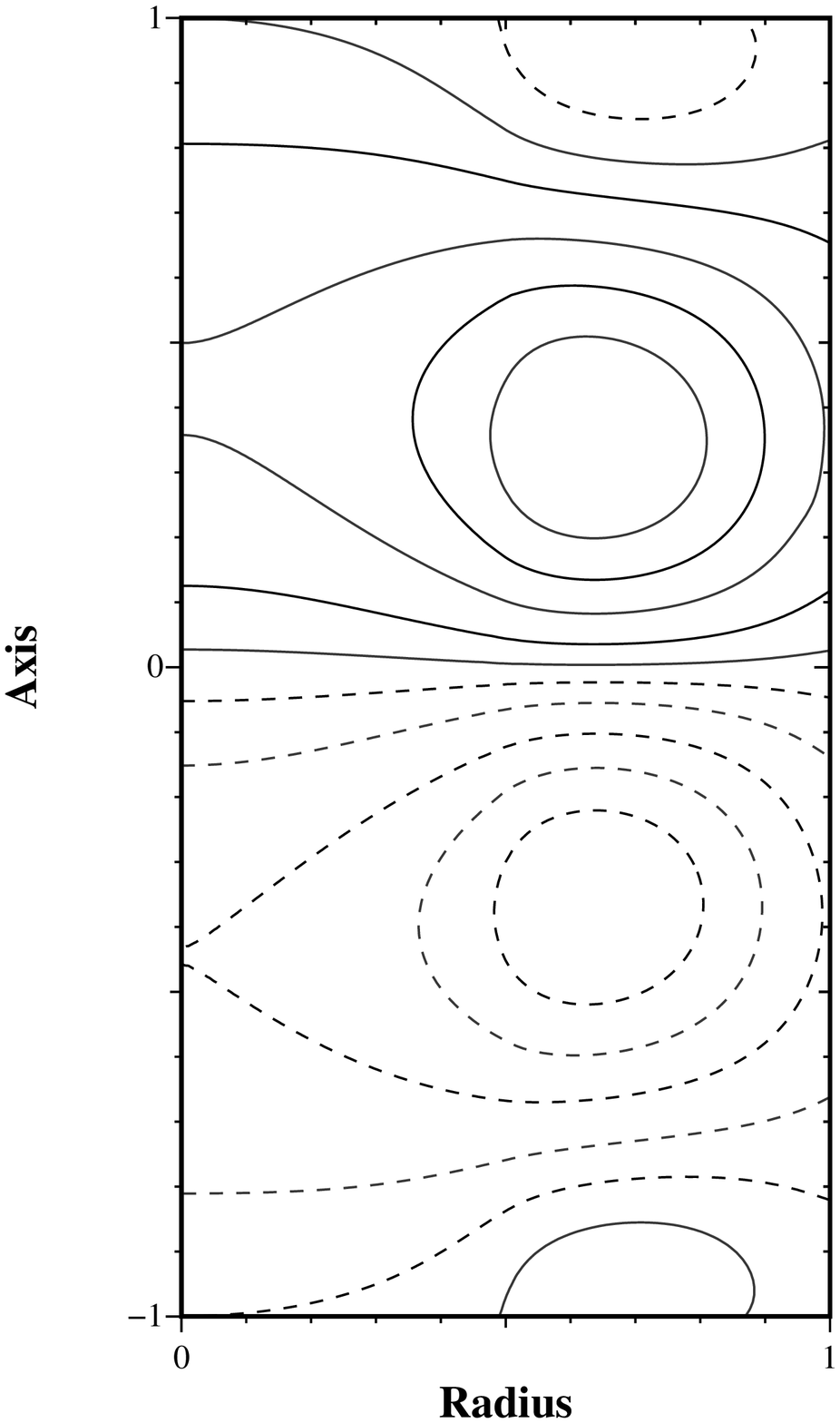,width=0.15\textwidth,bb=130
      76 451 683, clip}
    \epsfig{file=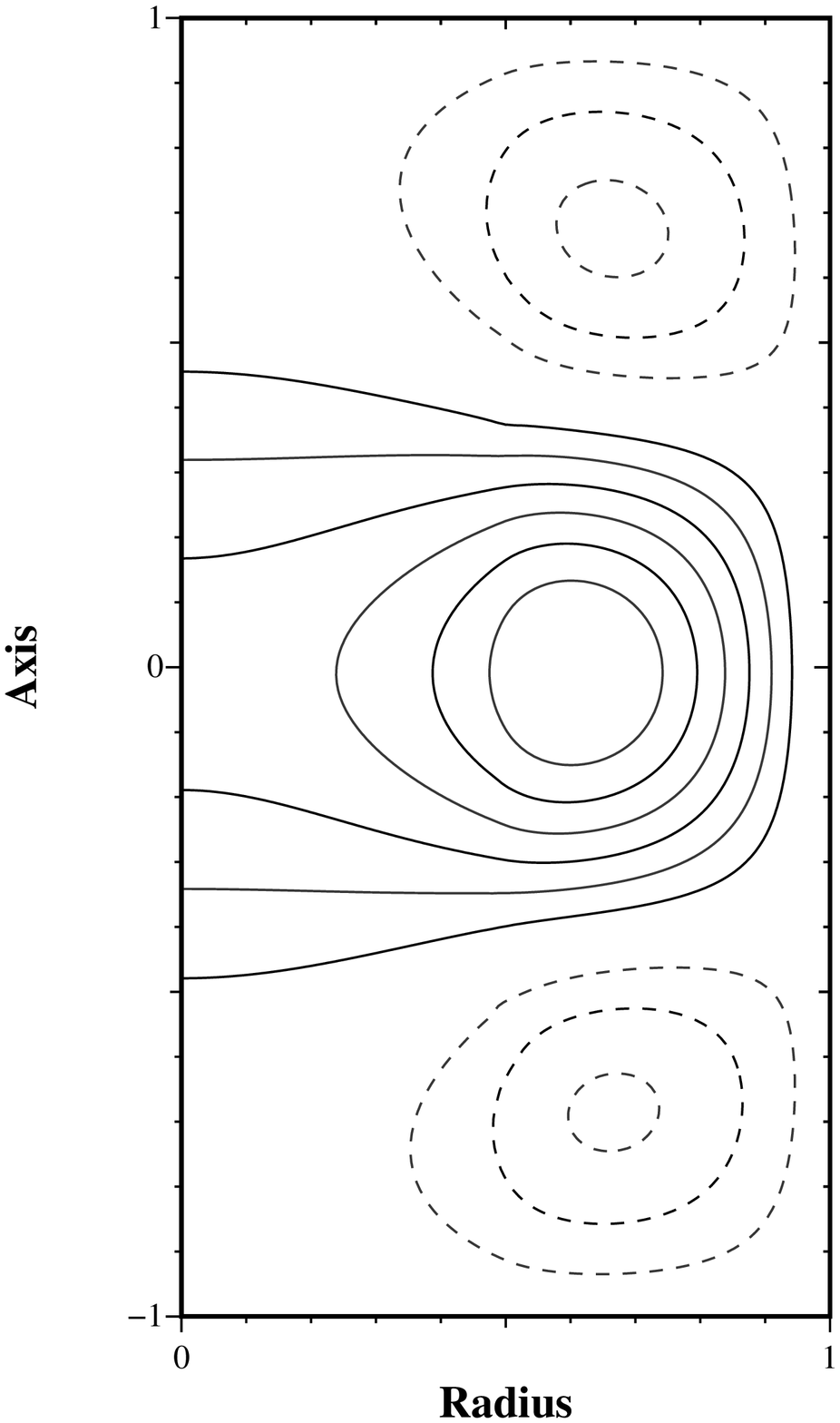,width=0.15\textwidth,bb=130
      76 451 683, clip}
    \epsfig{file=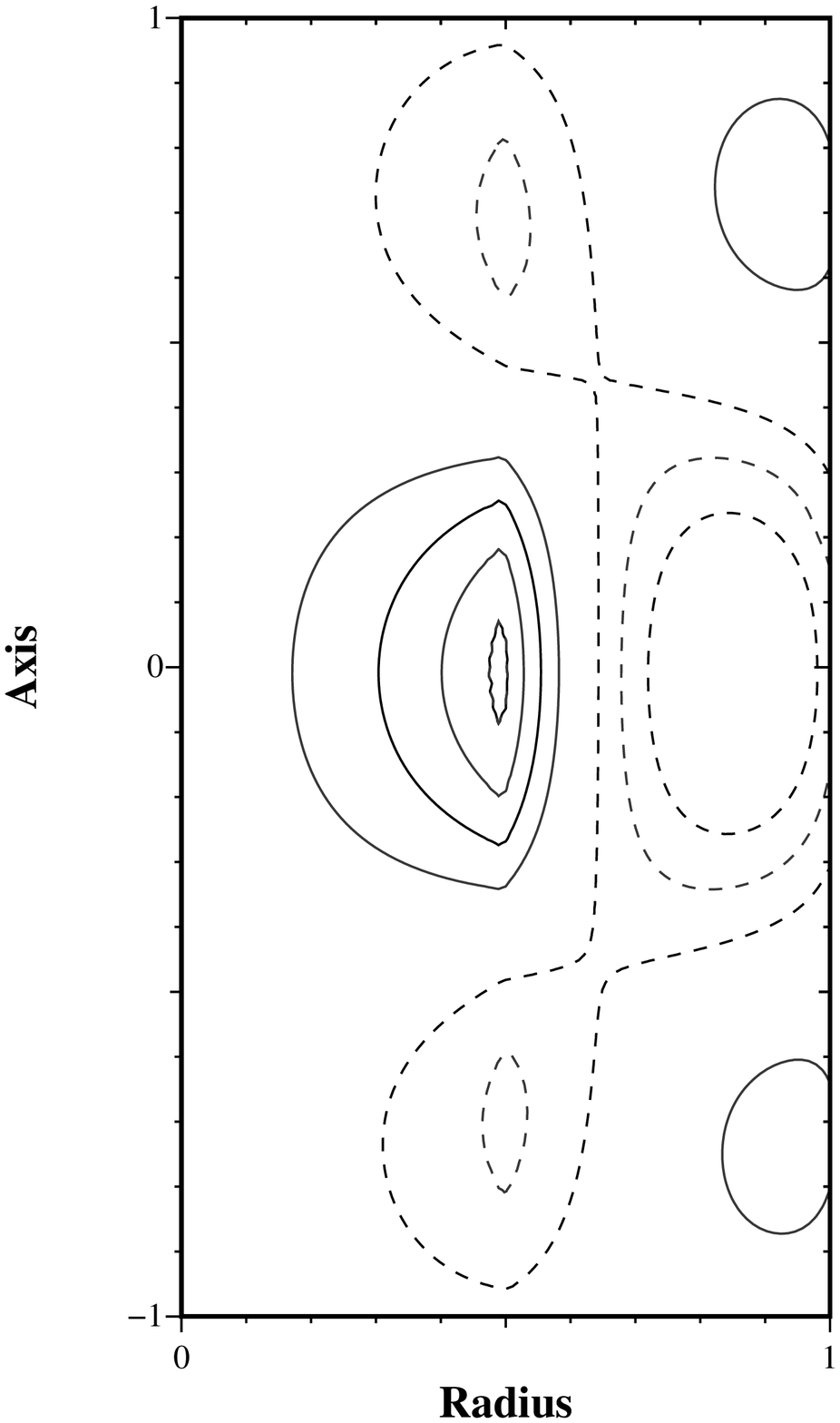,width=0.15\textwidth,bb=130
      76 451 683, clip} } \centerline{(a) \hfil \hfil (b) \hfil \hfil
    (c) \hfil \hfil \hfil \hfil \hfil (a') \hfil \hfil (b') \hfil
    \hfil (c') \hfil}
  \caption{ Geometric structure of the magnetic field of the
    $\alpha^2$-dynamo in a finite cylinder of aspect ratio $H/R=2$
    with an annular $\alpha$-effect: (a-c) $(m=0)$,
    $C_{\alpha}^{\rm{crit}}=20.2$ and $R_i/R=0.8$ and for (a'-c')
    $(m=1)$, $C_{\alpha}^{\rm{crit}}=8.8$ and $R_i/R = 0.5$ Represented
    are the radial (a, a'), azimuthal (b, b') and axial (c, c')
    components. Data results from the SFEMaNS approach.}
\label{fig:EsR_finite_08}
\end{figure}

We compare the critical dynamos numbers obtained in periodic and
finite cylinders for the same aspect ratio $H/R= 2$.  We compare three
cases: $R_i/R=0$, $0.5$, $0.8$.  The results are summarized in
Table~\ref{tab:perio_K}. All the eigenmodes are steady at the
threshold, the bifurcation is therefore of pitchfork type.
\begin{table}
\tbl{Periodic and finite cylinders of aspect ratio $H/R=2$ and
  different annular distributions of $\alpha$. The value $7.15^*$
  results from unpublished data from \citet{azp..2009}.}
{\begin{tabular}{|r|r|r|r|r|}
  \hline
  Case & Periodic & Periodic & Finite  & Finite\\
  \hline
  $R_i/R$ & $C_{\alpha}^{\rm{crit}}(m=0)$ &$C_{\alpha}^{\rm{crit}}(m=1)$ & $C_{\alpha}^{\rm{crit}}(m=0$) &$C_{\alpha}^{\rm{crit}}(m=1)$\\
  \hline
  0 & $7.15^*$ & 4.8 & 7.5 & 5.2\\
  0.5 & 9 & 8 & 9.4 & 8.8\\
  0.8& 20 & 20.2 & 20.2 & 20.4\\
  \hline
\end{tabular}}
\label{tab:perio_K}
\end{table}
Table~\ref{tab:perio_K} shows that, depending on the value of $R_i/R$,
there is a change of structure of the critical mode. In the periodic case
the mode $(m=0)$ is critical when $R_i/R=0.8$ with
$C_{\alpha}^{\rm{crit}}=20$ and the axial wavelength is $\lambda = 1$.
The mode $(m=1)$ is critical when $R_i/R=0.5$ with
$C_{\alpha}^{\rm{crit}}=8$ and the axial wavelength is $\lambda = 2$.
The corresponding magnetic eigen-modes are shown in
Figure~\ref{fig:EsR_perio}.  Note that, as expected, the magnetic
field is concentrated in the region where $\alpha$ is nonzero.
It is striking that the finite and the periodic results are similar when
the ratio $R_i/R$ is nonzero (see Table~\ref{tab:perio_K} and
Figure~\ref{fig:EsR_finite_08}). 
The only
notable effect of the top and bottom boundary conditions is to
concentrate the fields inside the box.  
For example, the structure of
the mode $(m=0)$ shown in Figure~\ref{fig:EsR_finite_08}(a-c) is
concentrated around the equator and is dominated by a quadrupolar
stationary state like a 's2t2' state.

The main conclusion of this section is that axisymmetric
$\alpha^2$-dynamos can be obtained in cylindrical geometries. 
In a finite cylinder with uniform distribution of $\alpha$ a
dominant axisymmetric mode occurs when the aspect ratio is below the
critical value $H/R\approx 0.79$.  For $H/R=2$, an
annular distribution of $\alpha$ can also generate an axisymmetric magnetic field if
the $\alpha$-effect is sufficiently localized. 
Note that \citet{2004GApFD..98..225T} obtained also an axial dipole (in a
sphere) using an anisotropic magnetic
diffusivity ($\eta_{\parallel} > \eta_{\perp}$).
Compared to similar studies in spherical geometry, we have explored only a small region of
the parameter space such as the spatial distribution of $\alpha$, its
symmetry properties, and its tensor structure.  This study has
validated our two different codes since they obtained identical
results on simple configurations.  This gives us some confidence
before turning our attention to the VKS experiment.

\section{Mean field dynamos with a VKS type flow}
\label{sec:VKS_alpha}
\subsection{Experimental setup and mean (axisymmetric) velocity field}
\begin{figure}
  \centering
  \includegraphics[width=0.5\textwidth]{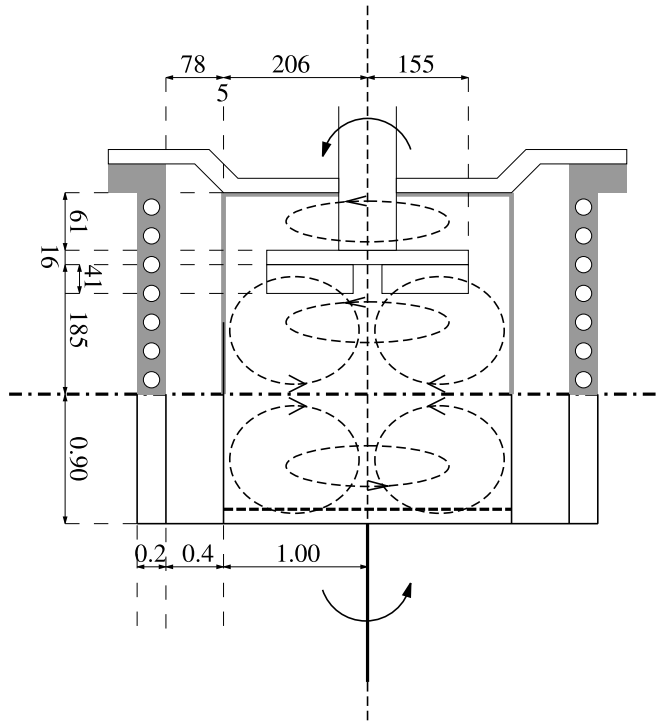}
  \caption{\label{fig2a} VKS set-up design and mean-flow structure:
    (upper part) dimensions and technical details of the experimental set-up, with the
    copper vessel including the cooling system, the thin envelope
    separating the flow and the stagnant liquid sodium, the impeller
    with fitted blades and the shaft. Courtesy of the VKS team.
    (lower part) Simplified geometry used in the numerical simulations in
    non-dimensional units. The conductivity jump between the liquid
    sodium and the copper vessel is taken into account.
The dashed line in the lower half denotes the region where dynamo action is
supported by the localized $\alpha$-effect 
(see section~\S~\ref{sec:localized_alpha}). Note, that the simulations do not
consider the region behind the impellers (lid layers).}
\label{fig::sketch}
\end{figure}

In the von-K\'arm\'an-Sodium  experiment a flow of liquid sodium
is driven by two counter-rotating impellers located at the opposite
ends of a cylindrical vessel.  Self-generation of a magnetic field
occurs if the magnetic Reynolds number exceeds the critical value
${\rm{Rm}^{\rm{crit}}}\approx 32$.  It is important to note though
that the dynamo is activated only when impellers made of soft iron are
used. A sketch of the experimental setup is shown in
Figure~\ref{fig::sketch}.
The vessel is filled with liquid sodium.  Two counter-rotating
impellers are located at the endplates of the vessel each fitted with eight
bended blades. 
The flow generated by the rotating blades is of von-K\'arm\'an type
and has been extensively investigated in \citet{2005PhFl...17k7104R}.
This so called s2t2-flow basically consists of two poloidal and two
toroidal cells.  

\begin{figure}
\centering
\includegraphics[width=0.5\textwidth,bb=67 359 468 696,clip=true]{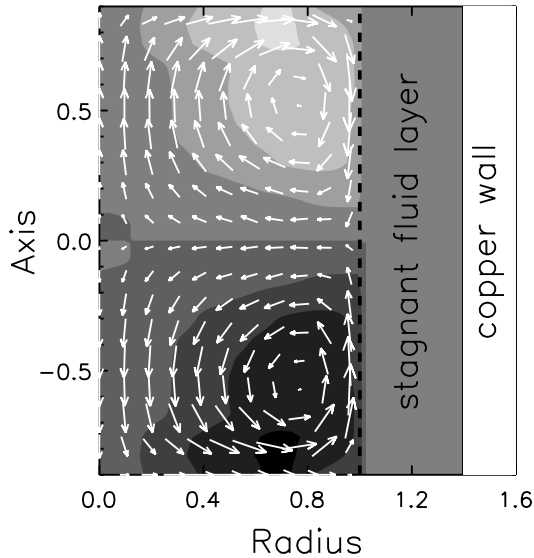}
\caption{\label{fig2b} Axisymmetric velocity field obtained from water
  experiments performed by \citet{2005PhFl...17k7104R} and which is applied in present kinematic simulations. Grey scale
  coded structures represent the azimuthal velocity field $u_{\theta}$ and the arrows represent
  the poloidal components ($u_r, u_z$).}%
\label{plot::axisym_vel_field}
\end{figure}

Throughout this paper, units are non-dimensionalized by setting the radius of
the flow domain to $R=1$
(corresponding to 0.206~m in the experiment, see Fig.~\ref{fig::sketch}) and
the magnetic diffusivity to $\eta=1$ (corresponding to
$\eta_{\rm{Na}}\approx 0.1~{\rm{m}}^2/{\rm{s}}$ in the experiment). Dimensional
values for the (maximum of the) velocity or the magnitude of the
$\alpha$-effect are then obtained in units of ${\rm{m}}{\rm{s}}^{-1}$ by a
multiplication with a factor close to $0.5$.
To facilitate the dynamo action the flow region is
surrounded by a layer of stagnant sodium of thickness 0.4 which is
enclosed by a solid wall of copper of thickness 0.2. 
The fluid velocity is zero when $r>R=1$. The copper wall is modeled by a
high conductivity region, i.e.  $\sigma_{\rm{Cu}}\approx 4.5
\sigma_{\rm{Na}}$ when $1.4 \le r \le 1.6$.
The flow is characterized by the magnetic Reynolds number
\begin{equation}
{\rm{Rm}}=\frac{{U_{\rm{max}} R }}{\eta}
\label{eq::def_magReynolds}
\end{equation}
where $U_{\rm{max}}$ is
the maximum of the fluid velocity.
A typical example for a flow field realized in a water experiment in
the VKS geometry is shown in
Figure~\ref{plot::axisym_vel_field}. It is time-averaged, axisymmetric
and symmetrized with respect to the
equator. We utilize this mean velocity field in the kinematic induction equation
\eqref{eq::inductioneq}.

\begin{figure}
\centering
\includegraphics[width=0.5\textwidth,bb=64 60 530
429,clip=true]{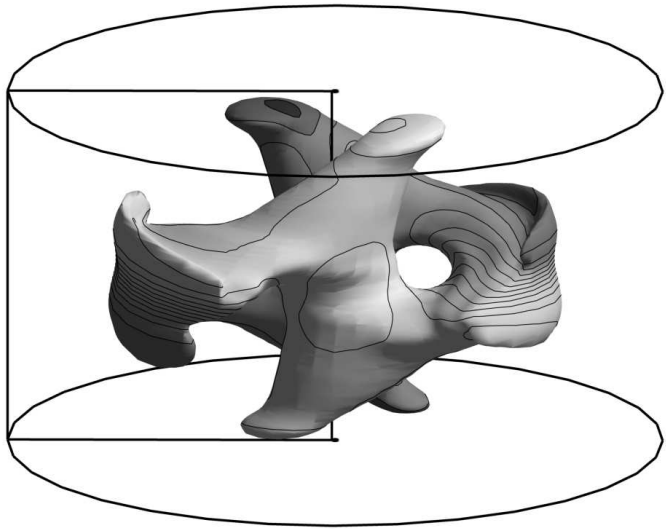}
\caption{\label{fig2cc} Iso-surface of the
  magnetic energy density at 40\% of the maximum value. The grey scale represents the azimuthal
  component $B_{\theta}$.}%
\label{plot::isosurf_bfield_noalpha}
\end{figure}

Using our two codes, we find that the critical Reynolds number for the mode
$(m=1)$ for this setting is approximately
${\rm{Rm}}^{\rm{crit}}\approx 45$ without any $\alpha$-effect. The
geometric structure of the eigenmode obtained for $\rm{Rm}=60$ is
shown in Figure~\ref{plot::isosurf_bfield_noalpha}. The picture shows
the iso-surface of the magnetic energy density $(2\mu_0)^{-1}|\vec{B}|^2$
at 40\% of the maximum value. The grey scaled map on the
iso-surface represents the azimuthal component of the magnetic
induction. This map shows that the mode $(m=1)$ dominates.  The well
known embracing banana-like structure is the dominating feature in the
central region.  The accumulation of magnetic energy close to the
equator at the interface between the flow domain and the surrounding
stagnant layer of fluid is another striking feature.
Because of the equatorial symmetrization of the velocity field, the resulting eigenmode does not exhibit any
azimuthal drift and remains stationary \citep{2003EPJB...33..469M}.

\subsection{$\alpha$-model}
\label{sec:alpha_model}

\begin{figure}
\centering
\includegraphics[width=0.9\textwidth,bb=77 144 994 442,clip=TRUE]{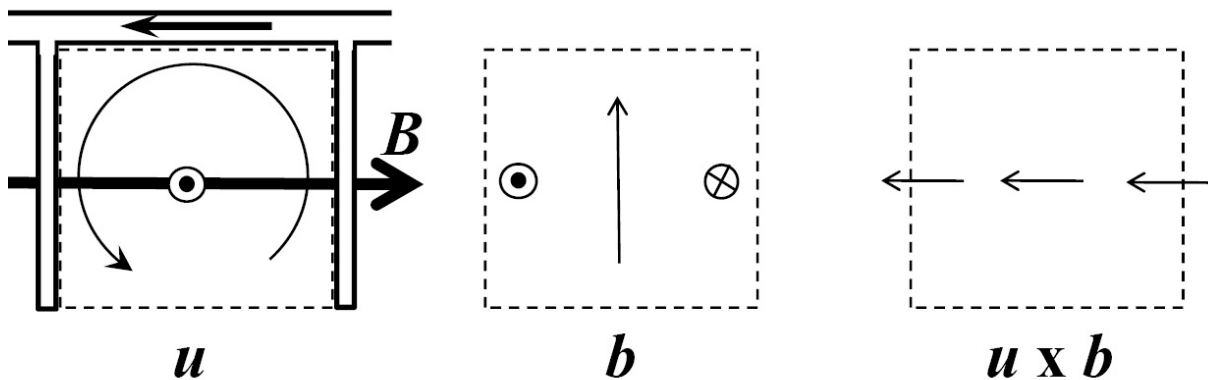}%
\caption{Sketch of the $\alpha$-model in the VKS experiment: between
  two impeller blades, (left) the fluid velocity (thin arrow and
  outwards symbol) acting on a given azimuthal mean magnetic field
  $\vec{B}$ (thick arrow) results in a secondary magnetic field
  $\vec{b}$ (middle) and generates an electromotive force $\vec{u}
  \times \vec{b}$ (right).}
\label{plot::alpha_vks_sketch}
\end{figure}

We now want to explore the implications of the suggestions that the strong shear between the flow trapped between the
impeller blades and the slower rotating fluid neighboring the
impeller results in a helical outward driven flow.  The induction
effects of such a flow can be parametrized by an $\alpha$-effect as
illustrated in Figure~\ref{plot::alpha_vks_sketch}.  In this sketch an
azimuthal electromotive force is produced from the cross-product of
the flow $\vec{u}$ between the blades and the magnetic field
$\vec{b}$ obtained from the distortion of an applied azimuthal
magnetic field. The resulting electromotive force is parallel to the
applied magnetic field but of opposite direction.  This corresponds to
a negative coefficient $\alpha_{\theta \theta}$ in the $\alpha$-tensor.
Applying a vertical magnetic field leads to a vertical electromotive
force again parallel to the applied magnetic field and of opposite
direction.  This corresponds to a negative coefficient $\alpha_{z z}$
in the $\alpha$-tensor.

In order to estimate the value of the coefficients $\alpha_{\theta
  \theta}$, $\alpha_{zz}$, we first estimate the magnetic Reynolds
number of the fluid flow trapped between the impeller blades.  Let
$U_{\rm{disk}}$ be the rotation speed at the rim of the disc. The maximum fluid velocity
  is obtained just below the upper impeller and has been measured to be $U_{\rm{max}} \sim \nu U_{\rm{disk}}$
with $\nu=0.6$ denoting an impeller efficiency
  parameter \citep{2005PhFl...17k7104R}. 
Then an estimate of the
recirculation flow intensity is $u\sim (U_{\rm{disk}}-U_{\rm{max}})/2$.  This leads to
$u/U_{\rm{disk}} \sim (1 - \nu)/2=0.2$.  The height of a blade is $h\approx
H/20\approx (1.8/20)R=0.09 R$, $R$ being the disc radius and $H$
the distance between the discs. 
Reminding the definition of the global magnetic Reynolds
 number from Eq.~(\ref{eq::def_magReynolds}), 
${\rm{Rm}}=U_{\rm{max}}R/\eta\approx \nu U_{\rm{disk}} R/\eta$ we define a
  local magnetic Reynolds number that characterizes the flow between the blades as 
${\rm{Rm}}_{\rm{blade}}\approx u h/\eta$.
The combination of both definitions yields
${\rm{Rm}}_{\rm{blade}}\approx (u/U_{\rm{disk}}) (h/R) ({\rm{Rm}}/\nu)$
leading to ${\rm{Rm}}_{\rm{blade}}\approx 1$ for
${\rm{Rm}}\approx 32$.  Therefore it seems reasonable to
apply a low magnetic Reynolds number approximation to estimate the value of the
$\alpha$-effect in the experiment according to the SOCA approximation~\citep{1980mfmd.book.....K}. 
This gives $\alpha_{\theta \theta}\sim
\alpha_{zz} \sim u u_r h / \eta$.  Assuming that the radial outward
flow $u_r$ is of the same order as $u$, for $\eta=0.1~{\rm{m}}^2{\rm{s}}^{-1}$ and
$R=0.2~{\rm{m}}$ we obtain $\alpha_{\theta \theta}\sim \alpha_{zz} \sim
5~{\rm{m}}{\rm{s}}^{-1}$. 
This estimate exceeds the value  ($\alpha\sim 1.8~{\rm{m}}{\rm{s}}^{-1}$) reported by
\citet{2008PhRvL.101j4501L} where the efficiency factor $\nu$ was not
considered in the definition of ${\rm{Rm}}$.  
Note, that this is only an evaluation of the order of magnitude
and, as the flow is highly turbulent, the optimum value of $\alpha$
estimated above probably is not realized in the experiment.

We henceforth focus our attention on $\alpha_{\theta\theta}$.  
It is indeed this coefficient that generates the poloidal component
of an axisymmetric magnetic field from its toroidal component.
In turn, the toroidal component is efficiently generated by twisting and
stretching the poloidal field component via a shear flow ($\omega$-effect).
The impact of an $\alpha_{zz}$ coefficient will be discussed at the end
of the next section.

\subsection{Localized $\alpha_{\theta\theta}$-effect results}
\label{sec:localized_alpha}
\begin{figure}\centering
\includegraphics[width=0.5\textwidth,bb=53 399 509 642,clip=true]{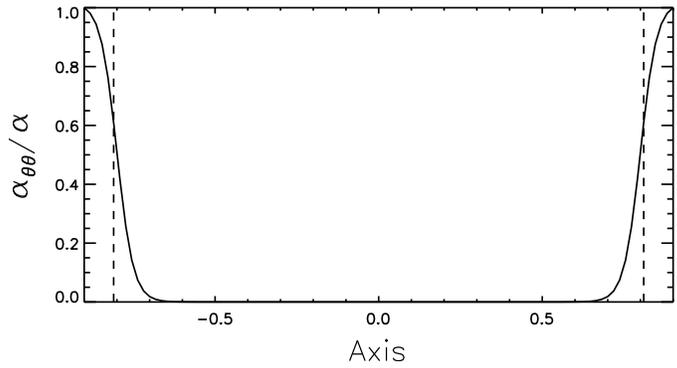}%
\caption{\label{fig2e} Axial distribution of the localized
  $\alpha$-effect. The magnitude of $\alpha$ is only significant
  in the impeller region indicated by the
  dashed lines.}%
\label{plot::axial_alpha}
\end{figure}

As the spatial distribution of $\alpha$ is not known a priori, we
start by performing simulations with the $\alpha$-effect
restricted to the impeller region. More specifically we set
\begin{equation}
  \alpha_{\theta\theta}(z)=
\begin{cases}
\alpha\left(1+\frac{1}{2}\left[\tanh\left(\frac{z-z_{\rm{top}}}{\Delta
          z}\right)-\tanh\left(\frac{z-z_{\rm{bot}}}{\Delta z}\right)\right]\right) 
& \text{if $0\le r\le R$}\\
0 
& \text{if $R\le r$}
\end{cases}
\label{eq::tanh}
\end{equation}
where $\alpha$ can be positive or negative, $\Delta z=0.05$,
$z_{\rm{top}}=z_{\rm{bot}}=\pm0.8$, and $R=1$ is the radial
extension of the impeller region.  Equation~\eqref{eq::tanh} specifies
a smooth transition of $\alpha_{\theta\theta}$ between a maximum value
$\alpha$ in the impeller region and a vanishing value in the bulk of
the container.  We henceforth refer to this model as the localized
$\alpha$-effect.  The axial profile of $\alpha_{\theta\theta}(z)$ is shown in
Figure~\ref{plot::axial_alpha}.

\begin{figure}
\centering
\includegraphics[width=0.7\textwidth,bb=60 357 541 702,clip=true]{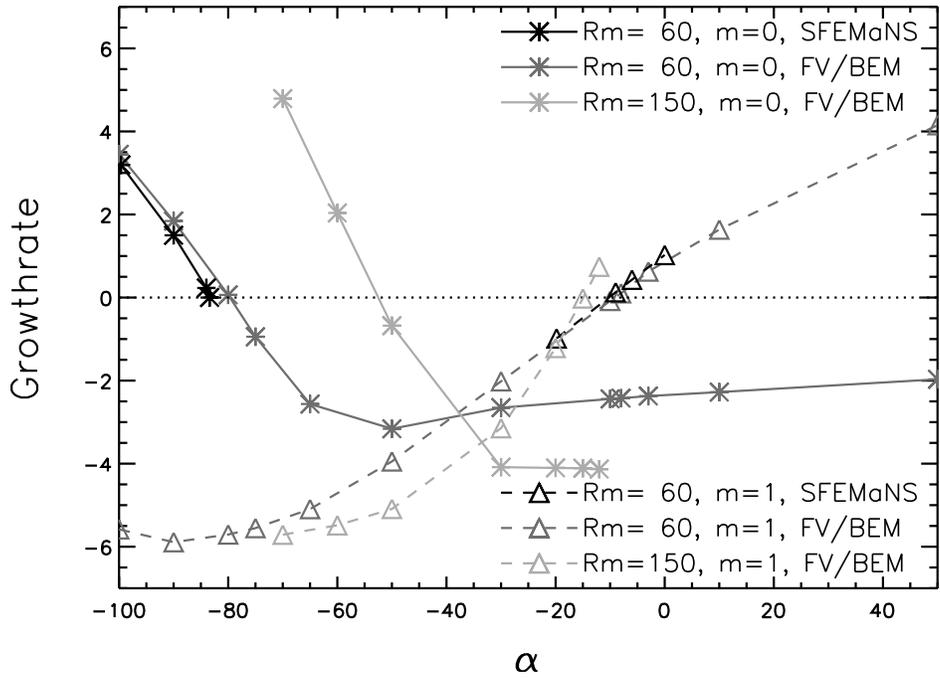}%
\caption{\label{fig2d} Field amplitude growthrates for the localized $\alpha$-effect.}%
\label{plot::gr_imp}
\end{figure}

\begin{figure}
\centering
\includegraphics[width=0.7\textwidth,bb=60 355 549
702,clip=true]{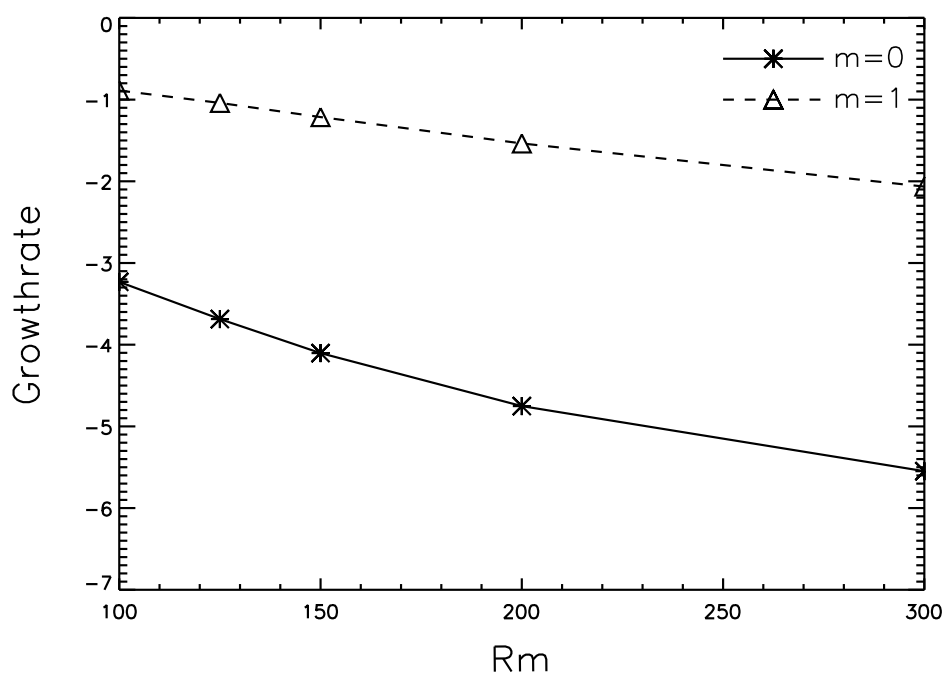}
\caption{\label{fig1777} Field amplitude growthrates for $\alpha=-20$ for the localized
  $\alpha$-distribution (\ref{eq::tanh}). The solid line denotes the axisymmetric mode
  $(m=0)$ and the dashed line denotes the mode $(m=1)$.}%
\label{plot::gr_vs_rm_a-20_imp}
\end{figure}

Figure~\ref{plot::gr_imp} shows the growthrates of the magnetic field
as a function of $\alpha$ for various magnetic Reynolds numbers.
The solid curves represent the growthrate for the axisymmetric mode
$(m=0)$ and the dashed curves represent the results for the mode
$(m=1)$.  We observe two distinct dynamo regimes that differ by the resulting field
geometry. The main characteristics of the dynamo solutions can be summarized as
follows.  For $\alpha \lesssim -50$ the eigenmode is axisymmetric
whereas for $\alpha \gtrsim -15$ the mode $(m=1)$ dominates.  We
observe in Figure~\ref{plot::gr_imp} that the growth rates of both
modes are negative in the range $[-50,-10]$.  This interval might
become smaller as ${\rm{Rm}}$ increases, but it seems that there
exists a non-empty range of $\alpha$ for which no dynamo action is
possible at all.  For instance, at $\alpha=-20$ no dynamo is obtained
for ${\rm{Rm}}$ up to $300$ and the corresponding growthrates even
decrease as ${\rm{Rm}}$ increases (see
Figure~\ref{plot::gr_vs_rm_a-20_imp}).

\begin{figure}
  \centering
  \includegraphics[width=0.7\textwidth,bb=60 357 541
  702,clip=true]{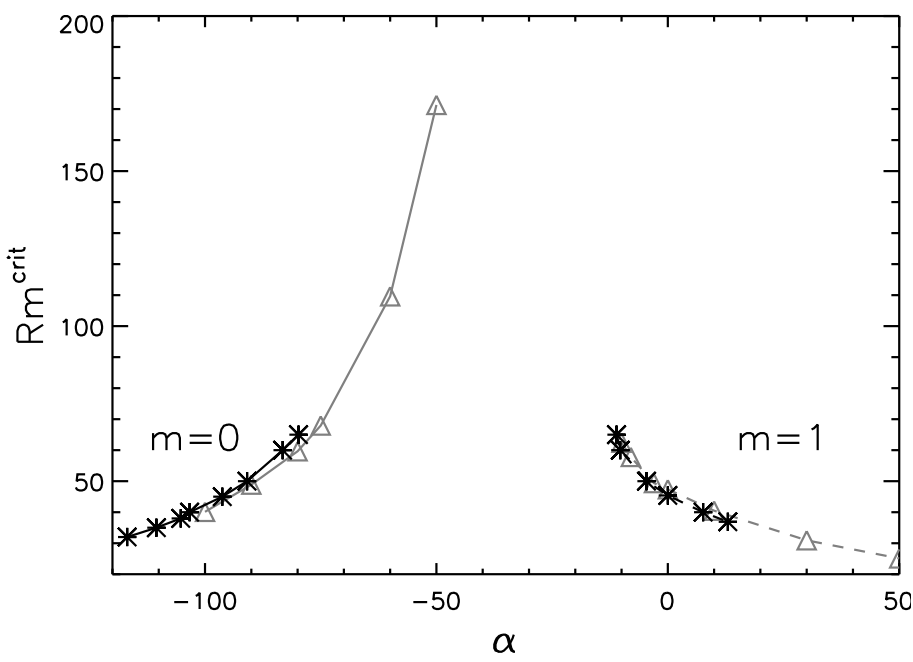}
  \caption{Critical magnetic Reynolds number in dependence of the
    magnitude of the $\alpha$-effect. The black curve represents the
    results obtained from the SFEMaNS scheme and the grey curve
    denotes the results of the FV/BEM approach. The left part (solid curves) shows
    ${\rm{Rm}}^{\rm{crit}}$ for the axisymmetric mode $(m=0)$ whereas
    the left part (dashed curves) is obtained for a dominating mode $(m=1)$.}%
\label{plot::rmcrit_vs_alpha_comp}
\end{figure}

The existence of
these two regimes becomes more obvious on
Figure~\ref{plot::rmcrit_vs_alpha_comp} where we show the critical
magnetic Reynolds number as a function of $\alpha$.
(${\rm{Rm}}^{\rm{crit}}$ is estimated by linear interpolation of two
adjacent growthrates obtained close to the dynamo threshold.)
The black curves represent the results provided by the SFEMaNS code
and the grey curves represent the results from the FV/BEM code.
Slight but systematic deviations between both codes are observed
especially for large negative values of $\alpha$.  Most probably the
disagreement is the result of a small difference in the localization
of $\alpha$ and/or the velocity field that is caused by the staggered
mesh definition in the FV/BEM scheme.  The present work does not
intend to perform a benchmark examination of different numerical
schemes so that the deviations are not investigated in detail as the
general trend and the critical values of the magnetic Reynolds number
are in rather good agreement.

By examination of the left branch of the graph in
Figure~\ref{plot::rmcrit_vs_alpha_comp} we estimate that
$\alpha_{\theta\theta}\approx -115$ is necessary to obtain an
axisymmetric dynamo at $\rm{Rm}\approx 32$.
In physical units this value corresponds to
$\alpha\approx 57.5~{\rm{m}}{\rm{s}}^{-1}$, which exceeds the estimated
magnitude of $\alpha$ by more than a factor of $10$.
Furthermore, this value is also four times larger than the maximum  fluid
velocity in the bulk of the cylinder. It is hard to believe that such a large
$\alpha$-effect is realized in the experiment.

We performed additional simulations with an $\alpha_{zz}$ coefficient located at the impellers.
The $\alpha$-tensor is therefore given by
$(0,\alpha_{\theta\theta},\alpha_{zz})$
with a tanh localized function (see Eq.~\ref{eq::tanh}) 
such that ${\alpha_{\theta\theta}}=\alpha_{zz}$.
We have found, for ${\rm{Rm=60}}$, $\alpha_c=-42.60$ with SFEMaNS and
$-44.6$ with FV/BEM.
These values are to be compared with $\alpha_c=-83.40$ for ${\rm{Rm}}=60$
obtained with a localized $(0,\alpha_{\theta\theta},0)$ $\alpha$-tensor. 
This reduction by a factor 2
is not enough for approaching realistic values of $\alpha$.

\subsection{Uniform $\alpha$-effect results}
In this section we assume that the $\alpha$-effect is uniformly
distributed in the flow region ($0 \le r \le 1, \, -0.9 \le z \le
0.9$). Here, only the $\alpha_{\theta\theta}$ component is assumed to
be nonzero.

\begin{figure}
  \centering
  \includegraphics[width=0.7\textwidth,bb=61 369 546
  703,clip=true]{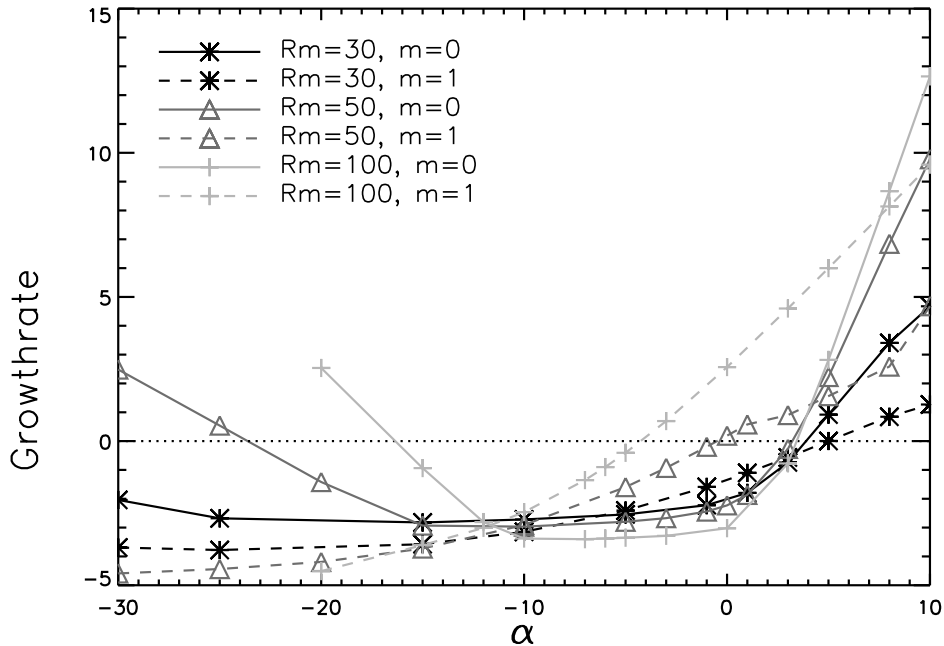}
  \caption{\label{fig1} Field amplitude growthrates for a homogenous
    $\alpha$-distribution for $\rm{Rm}=30, 50, 100$. The solid line
    denotes the axisymmetric mode $(m=0)$ and the dashed line denotes
    the mode $(m=1)$.}
  \label{plot::gr_vol}
\end{figure}

Figure~\ref{plot::gr_vol} shows the growthrates as a function of
$\alpha$ for ${\rm{Rm}}=30, 50 \mbox{ and } 100$.  These results have
been obtained with the FV/BEM code. We have verified on a few
simulations (not reported here) that the SFEMaNS code produces
similar results.

We see that for large negative values of $\alpha$ the mode $(m=1)$ is
suppressed in all the cases and the corresponding growthrates remain
negative.  The growthrate of the axisymmetric mode becomes positive
for sufficiently large magnetic Reynolds numbers.  Contrary to what is
observed when the $\alpha$-effect is localized, positive
growthrates of the axisymmetric mode are obtained for positive values
of $\alpha$.  For $\alpha \ga 0$ the growthrate becomes very sensitive
to small changes in $\alpha$, and a small interval exists
around $\alpha\approx +5$ where the critical magnetic Reynolds number
of the modes $(m=0)$ and $(m=1)$ are rather close together.  This
might be a promising possibility to obtain an axisymmetric field with
more reasonable values for $|\alpha|$.  This alternative needs to be examined
in more details because positive values for $\alpha$ can be obtained in the
equatorial layer where large scale intermittent vortices have been observed \citep{these..marie}.

\begin{figure}
  \centering
  \includegraphics[width=0.7\textwidth,bb=57 361 542
  702,clip=true]{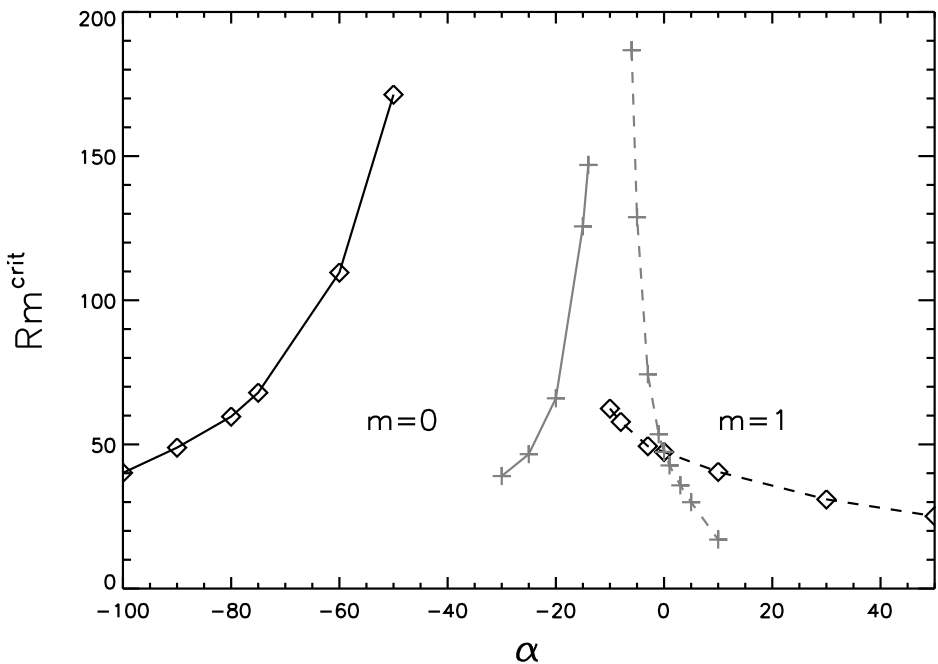}
  \caption{\label{fig6} ${\rm{Rm}}^{\rm{crit}}$ for different
    $\alpha$-distributions. Black curves: localized $\alpha$-effect. Grey
    curves: homogenous $\alpha$-effect. Solid lines denote the
    results for the mode ($m=0$), dashed lines denotes the results for $m=1$.}
  \label{plot::rmcrit_vs_alpha}
\end{figure}

For $-15\la\alpha\la -7$, again, there exists a region
that is characterized by an extremely high critical magnetic Reynolds
number, if a dynamo could exist at all.
Figure~\ref{plot::rmcrit_vs_alpha} shows the critical magnetic
Reynolds as a function of $\alpha$ for both the localized and the
uniform $\alpha$-effect. The
behavior of ${\rm{Rm}}^{\rm{crit}}$ for either $\alpha$-distributions
exhibits similarities like a strong tendency to diverge
around a certain value of $\alpha$.
Approximating the
curves with a power law ${\rm{Rm}}^{\rm{crit}}\propto
(\alpha-\alpha_s)^q$ gives a diverging behavior around
$\alpha_s\approx -18$ for the localized $\alpha$-effect and
$\alpha_{s}\approx -7.4$ for the homogenous distribution.

The critical range without dynamo action becomes significantly smaller
when the $\alpha$-effect is assumed to be uniform.
From the negative branch ($\alpha\la 0$, left graphs in
Figure~\ref{plot::rmcrit_vs_alpha}) we see that when the $\alpha$
distribution is uniform, the value of $|\alpha|$ which is necessary to
produce an axial dipole field at a certain $\rm{Rm}^{\rm{crit}}$ is
significantly lower than when the $\alpha$-effect is localized.  To
obtain an axisymmetric field at ${\rm{Rm}}^{\rm{crit}}\approx 32$ we
need to set $\alpha\approx -115$ when the distribution is localized
and $\alpha\approx-35$ when the distribution is uniform. Comparing the
volume fractions occupied by the $\alpha$-effect in both
configurations (which differ by a factor of about 10), it becomes
clear that the $\alpha$-effect within the impeller region operates
more effectively than in the remaining active zone.  However, even
when the $\alpha$-effect is assumed to uniformly penetrate the entire
flow domain the magnitude of $\alpha$ that is required to obtain the
mode $(m=0)$ at ${\rm{Rm}}^{\rm{crit}}\approx 32$ is still
$|\alpha|\approx |U_{\max}|\approx 16~{\rm{m}}{\rm{s}}^{-1}$ which remains
unrealistic high.

\section{Conclusion} \label{Sec:Conclusion} 
The present study is an illustration in the context of the fluid dynamo problem
of the interaction between numerical and experimental
approaches. An account of previous stages concerning the VKS experiment may be found
in~\cite{2008CRPhy...9..741L}.
We have shown that in the
framework of a mean-field model, axisymmetric $\alpha^2$-dynamos are
possible in cylinders embedded in vacuum. It seems always possible to
find a configuration (either determined from geometry or by a certain
spatial distribution of $\alpha$) which is able to generate an
axisymmetric magnetic field.  

In a VKS-like configuration, the
combination of an axisymmetric flow and an $\alpha$-effect can produce
axisymmetric magnetic modes as well. Our simulations with simple profiles of
$\alpha$ point out however that large and unrealistic values of $\alpha$
are necessary to explain the VKS experimental results. We could think of more complex
$\alpha$ distributions with negative values between the blades and positive values in
the equatorial layer. This is left for future work since a realistic assessment
would require to measure the kinetic helicity distribution in a water
model of the VKS device.

We note also that, even if realistic values of the magnitude of the
$\alpha$-effect had been obtained, one should have then to explain the
non-existence of the dynamo action when using steel impellers. Indeed, at
first sight, steel blades produce a flow, including kinetic helicity, close to the one
driven by the soft iron blades.  The role of the ferromagnetic
material to obtain the dynamo action appears to be a critical issue and
deserves further experimental and numerical investigations.

\section*{Acknowledgments}
We acknowledge fruitful discussions with R.~Avalos-Zu{\~n}iga.
We thank B. Knaepen and D. Carati of Universit\'e Libre de Bruxelles
for inviting some of us during the summer 2007 and the VKS Saclay team
for providing the mean flow used in section \S\ref{sec:VKS_alpha}.
This work was supported by ANR
project no. 06-BLAN-0363-01 ``HiSpeedPIV''. The computations
using SFEMaNS were carried
out on the IBM SP6 computer of Institut du D\'eveloppement et des
Ressources en Informatique Scientifique (IDRIS) (project \# 0254).
Financial support from Deutsche Forschungsgemeinschaft (DFG) in frame of the
Collaborative Research Center (SFB) 609 is gratefully acknowledged.

\appendix
{
\section{Numerical methods}
\subsection{SFEMaNS algorithm}
The SFEMaNS acronym stands for Spectral Finite Element method for Maxwell
and Navier-Stokes equations. This code is designed to solve the MHD equations
in axisymmetric domains in three space dimensions. To simplify the presentation 
we restrict ourselves in this Annex to the kinematic induction equation.

Let us consider a bounded domain $\Omega\subset \Real^3$ with boundary
$\Gamma=\partial\Omega$. The domain $\Omega$ is assumed to be partitioned into a
conducting region (subscript $c$) and an insulating region (subscript
$v$): $\overline{\Omega}=\overline{\Omega}_c\cup \overline{\Omega}_v,
\, \Omegac\cap\Omegav=\emptyset$.  The interface between the
conducting region and the nonconducting region is denoted by
$\Sigma=\partial\Omegac\cap\partial\Omegav$.  To easily refer to
boundary conditions, we introduce $\frontc=\front\cap\partial\Omegac,
\,\frontv=\front\cap\partial\Omegav$.  Note that
$\front=\frontv\cup\frontc$.

\subsubsection{The PDE setting}
\label{Elimination of the electric field}
The electromagnetic field
in the entire domain $\Omega = \Omegac \cup \Omegav$ 
is modeled by the Maxwell equations in the MHD limit. 
\begin{equation}
\left\{\begin{aligned}
&\mu\partial_t \bH = -\ROT\bE,\text{ in $\Omega$} \\
&\ROT\bH = 
\begin{cases}
\sigma (\bE + \vec{u} \CROSS \mu \bH) + \bj^s, & \text{in $\Omegac$} \cr
0,                                       & \text{in $\Omegav$}
\end{cases} \\
&\DIV\bE = 0 , \text{ in $\Omegav$} \\
&\bE\times\bn|_{\front} = \ba, \qquad \bH|_{t=0} =\bH_0, \text{ in $\Omegac$}
\end{aligned}\right. \label{dim_unsteady_degenerate_pb}
\end{equation} 
where $\bn$ is the outward unit normal on $\front$.
The independent variables are space and time.  The dependent variables
are the magnetic field, $\bH=\bB/\mu$, and the electric field, $\bE$.  The
data are the initial condition, $\bH_0$, the boundary data, $\ba$, and
the externally imposed distribution of current, $\bj^s$.  The data are
assumed to satisfy all the usual compatibility conditions, i.e. $\DIV
(\mu\bH_0)=0$.  The physical parameters are the magnetic permeability,
$\mu$, and the conductivity, $\sigma$. For the sake of generality, the
permeabilities in each domain ($\muc, \muv$) are distinguished, but we
take $\muc= \muv=\mu_0$ in the applications considered in the paper.

\subsubsection{Introduction of $\phi$ and elimination of $\bE$}
We assume
that the initial data $\bH_0$ is such that $\ROT \bH_0|_{\Omegav}=0$. We also
assume that either $\Omega_v$ is simply connected or there is some mechanism
that ensures that the circulation of $\bH$ along any path in the
insulating media is zero.  The condition $\ROT \bH|_{\Omegav}=0$ then
implies that there is a scalar potential $\phi$, defined up to an
arbitrary constant, such that $\bH|_{\Omegav}=\GRAD\phi$. Moreover, we
can also define $\phi_0$ such that $\bH_0|_{\Omegav}=\GRAD \phi_0$.
We now define
\begin{equation}
\bH=\begin{cases}
\bHc & \text{in $\Omegac$} \cr
\GRAD\phi & \text{in $\Omegav$},
\end{cases}\qquad
\mu = \begin{cases}\mu^c & \text{in $\Omegac$} \cr
\mu^v & \text{in $\Omegav$}, \end{cases}
\end{equation}
and we denote $\bn^c$ and $\bn^v$ the outward normal 
on $\partial\Omegac$ and  $\partial\Omegav$, respectively.
It is possible to eliminate the electric field from the problem and we finally obtain:
\begin{equation}
  \left\{\begin{aligned} &\muc\partial_t \bHc =
      -\ROT(\Rm^{-1}\sigma^{-1}(\ROT\bHc-\bj^s) - \vec{u}\CROSS\muc
      \bHc),
      &&\text{in $\Omegac$} \\
      &\muv \partial_t \LAP\phi = 0 &&\text{in $\Omegav$} \\
      &(\Rm^{-1} \sigma^{-1} (\ROT\bHc -\bj^s) - \vec{u}\CROSS\muc \bHc){\times}\bnc =
      \ba
      && \text{on $\frontc$} \\
      &\muv \partial_{{\bnv}}(\partial_t\phi) = -\bnv \SCAL\ROT
      (\bnv{\times}\ba),
      && \text{on $\frontv$} \\
      &\bHc{\times}\bnc +\GRAD\phi{\times}\bnv =0
      && \text{on $\Sigma$}\\
      &\muc\bHc\SCAL\bnc +\muv\GRAD\phi\SCAL\bnv =0
      && \text{on $\Sigma$}\\
      &\bHc|_{t=0} = \bHc_0, \qquad \phi|_{t=0} = \phi_0.
\end{aligned}\right. \label{unsteady_pb_H_phi_E_eliminated}
\end{equation}

\subsubsection{Weak formulation}

The weak formulation of \refp{unsteady_pb_H_phi_E_eliminated} that we want to
use has been derived in \cite{GLLN07}.  We introduce the following
spaces:
\begin{align}
\bL &= \{(\bbb,\varphi)\in \bL^2(\Omegac){\times} \tildeHunv\},\\
\bX &= \{(\bbb,\varphi)\in \Hrotc{\times}\tildeHunv;\ 
(\bbb{\times}\bnc + \GRAD\varphi{\times}\bnv)|_{\Sigma} = 0\}, \label{def_of_X}
\end{align}
and we equip $\bL$ and $\bX$ with the norm of $\bL^2(\Omegac){\times}\Hunv$ 
and $\Hrotc{\times}\Hunv$, respectively. 
$\tildeHunv$ is the subspace of $\Hunv$ composed of the functions of zero mean value.
The space $\Hrotc$ is composed of the vector-valued functions on $\Omega_c$
that are componentwise $L^2$-integrable and whose curl is also componentwise $L^2$-integrable.

We are now able to formulate the problem as follows:
Seek the pair $(\bHc,\phi)\in L^2((0,+\infty);\bX)\cap L^\infty((0,+\infty);\bL)$
(with $\partial_t \bHc$ and $\partial_t\phi$ in appropriate spaces) such that
for all $(\bbb, \varphi)\in \bX$ and $t\in (0,+\infty)$,
\begin{equation}
\left\{
\begin{aligned}
  &\bHc|_{t=0}=\bH^c_0;\quad \GRAD\phi|_{t=0}=\GRAD\phi_0,\\
  &\int_{\Omegac} \left[ \muc(\partial_t\bHc)\SCAL \bbb +
    ((\Rm\sigma)^{-1}(\ROT\bHc-\bj^s)-\vec{u}\CROSS\muc \bHc)\SCAL \ROT\bbb
  \right]
  +\int_{\Omegav} \muv(\partial_t\GRAD\phi)\SCAL \GRAD\varphi\\
  &\hspace{2cm} +\int_{\Sigma}((\Rm\sigma)^{-1}(\ROT\bHc-\bj^s)-\vec{u}\CROSS\muc \bHc)\SCAL
(\bbb\CROSS\bnc+\GRAD\varphi\CROSS\bnv)
\\
  &\hspace{2cm} =\int_{\frontc} (\ba{\times}\bn)\SCAL ( \bbb{\times}\bn) +
  \int_{\frontv} (\ba\CROSS\bn)\SCAL (\GRAD\varphi\CROSS\bn).
\end{aligned}
\right.
\label{eq_for_H_and_phi}
\end{equation}

The interface integral over $\Sigma$ is zero since
$\bbb\CROSS\bnc+\GRAD\varphi\CROSS\bnv=0$, but we nevertheless retain it
since it does not vanish when we construct the nonconforming
finite element approximation in
\S\ref{Sec:Finite_element_approximation}.

It has been shown in \cite{GLLN07} that \eqref{eq_for_H_and_phi} is
equivalent to \eqref{unsteady_pb_H_phi_E_eliminated}.  Observe that
the boundary conditions on $\frontv$ and $\frontc$ in
\refp{unsteady_pb_H_phi_E_eliminated} are enforced naturally in
\eqref{eq_for_H_and_phi}. The interface continuity condition
$\bHc{\times}\bnc +\GRAD\phi{\times}\bnv =0$ is an essential
condition, i.e. it is enforced in the space $\bX$, see
\refp{def_of_X}. One originality of the approximation technique
introduced in \cite{GLLN07} and recalled in
\S\ref{Sec:Finite_element_approximation} is to make this condition
natural by using an interior penalty technique. 

\subsubsection{Finite element approximation}
\label{Sec:Finite_element_approximation}
We approximate \eqref{eq_for_H_and_phi} by means of finite elements
in the meridian section and Fourier expansions in the azimuthal direction.

The generic form of approximations of $\bH^{c}$ and $\phi$ is
\begin{equation}
f(r,\theta,z,t) = \sum_{k=-M}^M f_h^k(r,z,t) e^{\text{i} k\theta},  \qquad 
\overline{f_h^k}(r,z,t) = f_h^{-k}(r,z,t), \ \forall k \in \overline{0,M}, 
\end{equation}
where $\text{i}^2=-1$ and $M+1$ is the maximum number of complex
Fourier modes. The coefficients $f_h^k(r,z,t)$ take values in finite
element spaces. We use quadratic Lagrange finite elements to
approximate the Fourier components of $\bHc$, i.e. the three
components of the magnetic field, $(H_r^c, H_\theta^c, H_z^c)$, are
continuous across the finite elements cells and are piecewise
quadratic. The approximation space for the magnetic field is denoted
$\bX_h^{\bH}$. Similarly, the Fourier components of the magnetic
potential are approximated with quadratic Lagrange finite elements.
The approximation space for the magnetic potential is denoted
$\bX_h^{\phi}$. No continuity constraint is enforced between 
members of $\bX_h^{\bH}$ and $\bX_h^{\phi}$.

The Maxwell equation is approximated by using the
technique introduced in \cite{GLLN07,GLLN09}. The main feature is that the
method is non-conforming, i.e. the continuity constraint
$(\bbb{\times}\bnc + \GRAD\varphi{\times}\bnv)|_{\Sigma} = 0$ in $\bX$
(see \refp{def_of_X}) is relaxed and enforced by means of an interior
penalty method.

We use the second-order Backward Difference Formula (BDF2) to
approximate the time derivatives. The nonlinear terms are made
explicit and approximated using second-order extrapolation in
time. Let $\Delta t$ be the time step and set $t^n:=n\Delta t$,
$n\ge 0$.
The solution to the Maxwell equation is computed by solving for $\bH^{c,n+1}$ in
$\bX_h^{\bH}$ and $\phi^{n+1}$ in $X_h^{\phi}$ so that the following
holds for all $\bbb$ in $\bX_h^{\bH}$ and all $\varphi$ in $X_h^{\phi}$
\begin{multline}
\int_{\Omegac} \left[ \muc\frac{D\bH^{c,n+1}}{\Delta t}\SCAL \bbb +
(\Rm\sigma)^{-1}\ROT\bH^{c,n+1}\SCAL \ROT\bbb \right] +\int_{\Omegac}
\muv\frac{D\phi^{n+1}}{\Delta t}\SCAL \GRAD\varphi\\
+\int_{\Sigma}((\Rm\sigma)^{-1}(\ROT\bH^{c,n+1}-\bj^s)-\vec{u}\CROSS\muc
\bH^*)\SCAL (\bbb\CROSS\bnc+\GRAD\varphi\CROSS\bnv)\\ +
g((\bH^{c,n+1},\phi^{n+1}), (\bbb,\varphi)) +s(\bH^{c,n+1},\bbb) \\ =
\int_{\Omegac} (\vec{u}{\times}\mu^c \bH^* + (\Rm\sigma)^{-1}
\bj^s)\SCAL \ROT\bbb +\int_{\frontc} (\ba{\times}\bn)\SCAL (
\bbb{\times}\bn) + \int_{\frontv} (\ba\CROSS\bn)\SCAL
(\GRAD\varphi\CROSS\bn),
\label{maxwell_step}
\end{multline}
where we have set
$D\bH^{c,n+1}:=\tfrac{1}{2}(3\bH^{c,n+1} -4\bH^{c,n}+\bH^{c,n-1})$,
$D\phi^{n+1}:=\tfrac{1}{2}(3\phi^{n+1} -4\phi^{n}+\phi^{n-1})$,
and 
\begin{align}
g((\bH^{c,n+1},\phi^{n+1}), (\bbb,\varphi))&:=\beta \sum_{F\in
\Sigma_h} h_F^{-1}\int_{F} (\bH^{c,n+1}\CROSS \bnc +
\GRAD\phi^{n+1}\CROSS \bnv)\SCAL (\bbb\CROSS \bnc + \GRAD\varphi\CROSS
\bnv),\\ s (\bH^{c,n+1},\bbb) &:= \gamma \int_{\Omegac}
\DIV(\muc\bH^{c,n+1}) \DIV(\muc\bbb).
\label{def_of_s}
\end{align}
The purpose of the bilinear form $g$ is to penalize the quantity
$\bH^{c,n+1}\CROSS \bnc + \GRAD\phi^{n+1}\CROSS \bnv$ across $\Sigma$
so that it goes to zero when the mesh-size goes to zero. The
coefficient $\beta$ is user-dependent. We usually take $\beta=1$. 
The purpose of the bilinear form $s$ is to have a control on the
divergence of $\bHc$.

\subsection{3D finite-volume/boundary-element-method (FV/BEM)}
\subsubsection{Finite volume method}
A finite volume (FV) approach 
provides a robust grid based scheme for the solution of the kinematic
induction equation.
The local spatial discretization on a regular grid is easy to implement
and allows a fast numerical solution of the induction equation in three dimensions.  
Physical quantities are defined at distinct locations on grid cells that
are obtained from a regular subdivision of the computational domain.  
The components of the magnetic field ${B}_{ix, iy, iz}$ at a grid cell labeled
by $(ix, iy, iz)$ are defined at
the center of the cell faces and the values are
interpreted as the average of the magnetic field on the specific cell face. 
The field update at a time step $n+1$ requires the discretization of Faraday's
law $\partial_t \vec{B}=-\nabla\times\vec{E}$.
This implies the computation of
the electric field ${E}_{ix,iy,iz}$ which is defined on the edges of a grid
cell so that the localizations of the components 
of $\vec{E}$ are slightly displaced with regard to the components of
$\vec{B}$.
Additional computational efforts occur as 
the computation of $\vec{E}$ requires the reconstruction of $\vec{u}$ and
$\vec{B}$ on the edge of a grid cell.
For moderate magnetic Reynolds numbers it is sufficient 
to apply a simple arithmetic average to interpolate $\vec{u}$ and
$\vec{B}$ on the edge.  
For larger magnetic Reynolds numbers, however, more elaborate schemes have to
be applied \citep[e.g.][]{2004JCoPh.196..393Z,2006JCoPh.218...44T}.   
The complications that arise by the definition of a second, staggered mesh are
essentially outweighed by the maintenance of the divergence free condition
and the conservation of the fluxes across interfaces between neighboring cells
which are intrinsic properties of the specific finite volume approach.

To relax the constraints of the time step restriction an implicit solver for the diffusive
part $ - \nabla\times(\eta\nabla\times\vec{B})$ 
of the induction equation is applied. 
The full scheme for the semi-implicit field update at time step $n+1$
is second-order in time and is summarized by the following expression
\citep{1999IJNMF..30..335K}:
\begin{equation}
\vec{B}^{n+1}=\vec{B}^n+\Delta t F^{\rm{exp}}\left[\vec{B}^n+\frac{\Delta t}{2}
  F[\vec{B}^n]\right]+\frac{\Delta
  t}{2}(F^{\rm{imp}}[\vec{B}^n]+F^{\rm{imp}}[\vec{B}^{n+1}]).\label{eq::semi_impl} 
\end{equation}
Here $F^{\rm{exp}}$ denotes the discretized operator accounting for the
terms of the induction equation that have been made explicit 
($\propto\vec{u}\times\vec{B}$ and $\alpha\vec{B}$),
$F^{\rm{imp}}$  denotes the discretized operator accounting for terms made implicit
 ($\propto\eta\nabla\times\vec{B}$) and $F=F^{\rm{imp}}+F^{\rm{exp}}$.  
\subsubsection{Insulating boundary conditions}
Insulating domains are characterized by a vanishing current 
$\vec{j}\propto \nabla\times\vec{B}=0$ so that, assuming that the vacuum domain
is simply connected,
 $\vec{B}$ can be expressed as
the gradient of a scalar magnetic potential
$\vec{B}=-\nabla\varPhi$. The potential $\varPhi$ is a solution to the Laplace equation
\begin{equation}
\Delta\varPhi =0, \quad
\varPhi \rightarrow O(r^{-2}) \mbox{ for } r\rightarrow\infty.
\label{eq::laplace}
\end{equation}
The computation of $\vec{B}$ at the boundary 
requires the integration of~$\Delta\varPhi~=~0$.
An effective approach to compute the potential $\varPhi$ and the corresponding
boundary field for insulating boundary conditions is provided by the boundary
element method (BEM). 
This procedure has been proposed in \citet{2004JCoPh.197..540I} and
\citet{2005GApFD..99..481I}, and was recently modified and applied to various
dynamo problems \citep{2008giesecke_maghyd}.  
We now give a short description of the technique.

Consider a volume $\Omega$ that is bounded by the surface 
$\Gamma$ and let ${\partial}/{\partial n}=\vec{n}\cdot\nabla$ denote 
the outward normal derivative.
After applying Green's second theorem including some straightforward
manipulations, the magnetic potential $\varPhi$ is shown to satisfy
the following integral equation:
\begin{equation}
\frac{1}{2}\varPhi(\vec{r})=\int\limits_{\Gamma} G(\vec{r},
\vec{r}')\underbrace{\frac{\partial \varPhi(\vec{r}')}{\partial
  n}}_{\displaystyle -B^{\rm{n}}(\vec{r}')}-\varPhi(\vec{r}')\frac{\partial
  G(\vec{r},\vec{r}')}{\partial n}d\Gamma(\vec{r}'), \label{eq::bie_phi}
\end{equation}
where $G(\vec{r}, \vec{r}')$ is the Green's function, or fundamental solution,
which fulfills $\Delta G(\vec{r},\vec{r}')=-\delta(\vec{r}-\vec{r}')$ and is
explicitly given by $G(\vec{r},\vec{r}')=-(4\pi\left|\vec{r}-\vec{r}'\right|)^{-1}$.
$\partial_n\Phi=-B^{\rm{n}}$ yields the normal component of
$\vec{B}$ on $d\Gamma$ which is known from the finite volume method as
described in the previous paragraph. 
The tangent components of the magnetic field  $B^{\tau}$ at the boundary
are computed by:
\begin{equation}
{B}^{\tau}\!=\!\vec{e}_{\tau}\!\cdot\!\vec{B}\!=-\vec{e}_{\tau}\!\cdot\!\nabla\!\varPhi(\vec{r}) 
\!=\!2\!\!\int\limits_{\Gamma}\!\!\vec{e}_{\tau}\!\cdot\!
\left(\!\varPhi(\vec{r}')\nabla_{r}\!\frac{\partial G(\vec{r},\vec{r}')}
{\partial n}\!+\!B^{\rm{n}}(\vec{r}')\nabla_{r}\!G(\vec{r},\vec{r}')\!\right)\!d\Gamma(\vec{r}')
\label{eq::bie_b}
\end{equation}
where $\vec{e}_{\tau}$ represents a tangent unit vector on the surface
element $d\Gamma(\vec{r}')$.
It can been shown that Eqs.~(\ref{eq::bie_phi}) and~(\ref{eq::bie_b})
contain all the contributions if all the field sources are located within the volume
enclosed by $\Gamma$.

The discretization of Eq.~(\ref{eq::bie_phi}) 
and Eq.~(\ref{eq::bie_b}) leads to an
algebraic system of equations which finally determines $B^{\tau}$ at a
single point on the surface by performing a matrix multiplication that connects
the normal components $B^{\rm{n}}$ 
at every surface grid cell of the computational domain:
\begin{equation}
\vec{B}^{\tau}_i=\sum_{i=1}^N\mathcal{M}_{ij}\vec{B}^{\rm{n}}_j.
\label{eq::matrixeq}
\end{equation}
Eq.~(\ref{eq::matrixeq}) introduces a global ordering of the quantities
${B}^{\tau}$ and $B^{\rm{n}}$ 
defined by an explicit mapping of the boundary grid-cell
indices $(ix, iy, iz)$ on a global index $i=0,1,2,\cdots,N$
where $N=2\cdot(nz\cdot ny+nz\cdot nx+ny\cdot nx)$ represents the total
number of boundary elements. 
The matrix elements $\mathcal{M}_{ij}$  are computed numerically applying a
standard 2D-Gauss-Legendre Quadrature method.
In general the matrix $\mathcal{M}$ is dense and requires a large amount of
computational resources. However, $\mathcal{M}$ only depends on the geometry
of the problem and therefore has only to be computed once. 
}
%
\bibliographystyle{gGAF} 
\bibliography{giesecke_AxisymMFDynVKS}

\label{lastpage}

\end{document}